\documentclass[twocolumn,prd,nofootinbib,longbibliography]{revtex4-1}

\newcommand{\beq}{\begin{equation}}
\newcommand{\eeq}{\end{equation}}

\def\mnras{M.N.R.A.S.}
\def\apjs{Astrophys.\ J.\ Supp.}
\def\apjl{Astrophys.\ J.\ Lett.}
\def\aap{Astron.\ Astrophys.}

\def\jcap{J.\ Cosm. Astroparticle Phys.}

\usepackage{graphicx}
\usepackage{amsmath}
\usepackage{amssymb}

\begin{document}
\title{
Testing cosmological models with large-scale power modulation using 
microwave background polarization observations
}
\author{Emory F. Bunn}
\email{ebunn@richmond.edu}
\author{Qingyang Xue}
\affiliation{Physics Department, University of Richmond, Richmond, VA  23173}
\author{Haoxuan Zheng}
\affiliation{Physics Department, Massachusetts Institute of
Technology, Cambridge, MA}

\begin{abstract}
We examine the degree to which observations of large-scale cosmic microwave background (CMB) polarization can shed light on the puzzling large-scale power modulation in maps of CMB anisotropy. We consider a phenomenological model in which the observed anomaly is caused by  modulation of large-scale primordial curvature perturbations, and calculate Fisher information and error forecasts for future polarization data, constrained by the existing CMB anisotropy data. Because a significant fraction of the available information is contained in correlations with the anomalous temperature data, it is essential to account for these constraints. We also present a systematic approach to finding a set of normal modes that maximize the available information, generalizing the well-known Karhunen-Lo\`eve transformation to take account of the constraints from the temperature data. A polarization map covering at least $\sim 60\%$ of the sky should be able to provide a $3\sigma$ detection of modulation at the level favored by the temperature data. A significant fraction of the information in such a data set is contained in the single mode that optimally encapsulates the signal due to temperature-polarization correlation.
\end{abstract}

\maketitle

\section{Introduction}

The cosmic microwave background (CMB) radiation provides much of the most important evidence in support of the standard cosmological model \cite{planck15overview,wmap9yrresults,bucherreview}. However, there have been claims of various ``anomalies'' on large angular scales in the all-sky maps made by the WMAP and Planck satellites, which appear to be in tension with certain aspects of the model \cite{wmapanomalies,planck13isotropy,planck15isotropy,schwarz04alignment,copi06,schwarz15}. Some, such as the alignment of low-order multipoles \cite{dOTZH,wmapanomalies,planck13isotropy,planck15isotropy,schwarz04alignment,copimv,landmagueijo,copi15alignment}
and the dipolar modulation of fluctuations \cite{wmapanomalies,planck13isotropy,planck15isotropy,eriksen04,eriksen07,gordon07,hansen09,hoftuft09,akrami14,adhikari15,aiola15}, even appear to violate the assumptions of homogeneity and isotropy. If there is indeed strong evidence that these assumptions are violated, the effect on cosmology would be revolutionary.

The statistical significance of these anomalies is controversial (e.g., 
\cite{wmapanomalies,efstathiou}), in large part because they are quantified with \textit{a posteriori} statistics -- that is, the anomalies were noticed in the data, and subsequently statistics were devised to quantify their improbability. Such statistics are problematic: in any large data set, some patterns will arise merely by chance, and statistics designed after the fact to characterize these patterns will have artificially low $p$-values. (This problem is sometimes described as the ``look-elsewhere effect.'') One might therefore choose to disregard the subject entirely. On the other hand, if the anomalies are not mere flukes, the consequences would be of the highest importance. It therefore seems reasonable to examine them closely while maintaining skepticism.

When faced with the problem of \textit{a posteriori} statistics, the natural solution is to seek a new data set for which the questions can be addressed \textit{a priori}. The large-angle CMB intensity has already been measured to the limit of cosmic variance, but CMB polarization on the largest angular scales have not yet been thoroughly characterized. In this paper, we examine the degree to which all-sky or partial-sky polarization data could help us to determine the significance of the dipolar modulation in fluctuation power.

Although measurements of CMB polarization have been made, there are none that have reliable information on the large angular scales of chief interest to us. We therefore do not use existing polarization data to constrain theories. (In constrast, see \cite{paci10,paci13}.) Rather, our focus is on the question of how much light \textit{future} polarization data, with reliable large-angle information, would shed on the modulation.

We focus specifically on the dipolar power modulation -- that is,
the observation that the fluctuations in one half of the sky appear
to be slightly larger in amplitude than in the other half.
We choose to examine this anomaly because it appears in some ways more robust than the others. In particular, when maps that have been filtered to contain non-overlapping multipole ranges are used to calculate the modulation direction, the results are remarkably consistent \cite{hansen09}. Under the hypothesis of statistical isotropy, these directions would be independent random variables. Even if we regard the first of these determinations as besmirched with the stain of \textit{a posteriori} statistics, the remainder are untainted.

Possible explanations for any of the anomalies come in three categories:
an anomaly can be a fluke, the result of a systematic error,
or a sign of new physics. In this paper, we disregard the possibility of systematic error, as the modulation appears robustly in different data sets
(WMAP and Planck), analyzed in different ways. We are therefore interested in the question of how well future polarization data could distinguish between the fluke hypothesis and the new-physics hypothesis. 

Under the fluke hypothesis, the Universe is described by the standard statistically-isotropic Gaussian model, and the probability distribution for polarization observations is straightforward to calculate. Note that the relevant probability distribution is the conditional probability for the future polarization data, constrained by the existing anisotropy data, as emphasized in \cite{copi13,yoho,odwyer}.

Previous work \cite{copi13,yoho,odwyer} has assessed the ability of 
polarization data to test the fluke hypothesis. In this paper,
we go further by comparing this hypothesis with
an alternative in which statistical isotropy is broken.
The most straightforward candidates for the new-physics hypothesis involve modulating the the primordial perturbations with a long-wavelength mode \cite{dvorkin,erickcek08,erickcek08b,erickcek09,moss,zibin,byrnes},
although there are other possibilities as well \cite{dai}.
Such a modulation can be produced in inflationary scenarios with a curvaton field, among other ways. Rather than committing to a specific physical model, 
we represent the new-physics hypothesis with a phenomenological model originally explored by Dvorkin et al. \cite{dvorkin}. We suppose that the primordial Newtonian curvature fluctuation
$\Phi$ is modulated by a multiplicative perturbation that breaks statistical
homogeneity and isotropy. To be specific, we suppose that
\beq
\Phi({\bf r})=g_1({\bf r})[1+h({\bf r})]+g_2({\bf r}).
\eeq
Here $g_1$ and $g_2$ are homogeneous and isotropic Gaussian random processes,
of the sort found in the standard model. The modulating field $h$ contains only very long-wavelength terms. In fact, to explain the dipolar modulation of the CMB, it can be taken to be a simple gradient 
\beq
h({\bf r})={{\bf w}\cdot
{\bf r}\over R_{LSS}}
\eeq
for some constant vector ${\bf w}$. 
Here $R_{LSS}$ is the distance to the last-scattering surface and is introduced to make ${\bf w}$ dimensionless.
Note that the magnitude of ${\bf w}$ differs from the parameter $w_1$ of ref. \cite{dvorkin} by a fixed normalization factor $w\equiv |{\bf w}| = 
\sqrt{3/(4\pi)}\ w_1$.

The reason for the two fields $g_1,g_2$ is that the modulation does not appear to persist to arbitrarily small length scales \cite{hirata09,hansonlewis}. As a result, we place the large-scale, modulated Fourier modes in $g_1$ and the small-scale, unmodulated modes in $g_2$. We adopt the simplest possible prescription: we let $g_1$ contain all Fourier modes below a fixed wavenumber cutoff $k_{\rm max}$, and place all modes with $k>k_{\rm max}$ in the unmodulated field $g_2$. The power spectra of $g_1$ and $g_2$ are taken to be the standard power spectrum $P(k)=Ak^{n-4}$ with spectral
index $n=1$, for $k<k_{\rm max}$ and $k>k_{\rm max}$ respectively.

We wish to quantify the new information that could be gained about this theory by a future polarization data set. Because CMB polarization is correlated with the temperature anisotropy, which has already been well-measured, we should consider the conditional probability distribution of the future polarization data, given the temperature anisotropy data we already have. In the theories under consideration, all of the probability distributions are Gaussian. 
To be specific, the joint probability distribution for temperature and polarization (whether expressed in position space or in the spherical harmonic basis) is a multivariate normal distribution with zero mean. The
conditional probability for polarization given temperature is then a normal distribution with nonzero mean. Both the mean and the covariance matrix of this distribution depend on the theory under consideration. To assess the ability of this data set to distinguish among competing theories, we will examine the theory-dependence of the distribution.

In particular, we will calculate the Fisher information for the parameter $w$, and show that for sufficiently large $k_{\rm max}$ a polarization data set could measure a $w$ value at the level suggested by the temperature data with $\sim 3\sigma$ significance, even with incomplete sky coverage. We will also show that a significant fraction of the information in such a data set is contained in a single mode, resulting from the correlation of the polarization with the known temperature data.

\section{Formalism}

\subsection{Constrained Gaussian random processes}

We begin by summarizing some results regarding constrained Gaussian random processes \cite{rybickipress,hoffman91,hoffman92,bunnwiener,lahav,bunnwiener2}. 
To be specific, we consider a set of data that can be modeled as a sample of a Gaussian random process with zero mean. Suppose that 
a subset of the data, represented by the vector
$\vec d_1$, has been measured, and that measurement of an additional data set $\vec d_2$ is planned. In the following sections of this paper, $\vec d_1$ will be the CMB intensity data already measured by Planck, and $\vec d_2$ will be a future set of polarization data. We combine the two data sets into a single vector 
\beq
\vec d_{\rm all}=\begin{pmatrix} \vec d_1\\ \vec d_2\end{pmatrix}
\eeq
The combined data is a sample from a Gaussian random process with mean zero and covariance matrix
\beq
\mathbf{M}_{\rm all}=\langle \vec d_{\rm all}\vec d_{\rm all}^T\rangle
=\begin{pmatrix}\mathbf{M}_{11} & \mathbf{M}_{21}^T \\ \mathbf{M}_{21} & \mathbf{M}_{22}\end{pmatrix},
\eeq
where $T$ denotes transpose and $\mathbf{M}_{jk}=\langle \vec d_j\vec d_k^T\rangle$.
The matrix $\mathbf{M}_{\rm all}$ depends on a set of theory parameters $\vec\theta$.
The likelihood function is
$p(\vec d_{\rm all}|\vec\theta)\propto e^{-\chi^2/2}$, with
\beq
\chi^2=\vec d_{\rm all}^T\mathbf{M}_{\rm all}^{-1}\vec d_{\rm all}.
\eeq
Because we have already measured $\vec d_1$, we wish to know
the conditional probability $p(\vec d_2| \vec d_1,\vec\theta)$.
This is still proportional to $e^{-\chi^2/2}$, but now we regard
$\vec d_1$ as fixed. 

Let 
\beq
\mathbf{M}_{\rm all}^{-1}=\mathbf{N}=
\begin{pmatrix}\mathbf{N}_{11} & \mathbf{N}_{21}^T \\ \mathbf{N}_{21} & \mathbf{N}_{22}\end{pmatrix}.
\eeq
Then
\begin{align}
\chi^2&=\vec d_1^T \mathbf{N}_{11}\vec d_1 
+\vec d_1^T \mathbf{N}_{21}^T \vec d_2 + \vec d_2^T \mathbf{N}_{21}\vec d_1+
\vec d_2^T \mathbf{N}_{22}\vec d_2\\
&=(\vec d_2-\vec\mu)^T \mathbf{N}_{22}(\vec d_2-\vec\mu)+\mbox{constant,}
\end{align}
where 
\beq
\vec\mu = -\mathbf{N}_{22}^{-1}\mathbf{N}_{21}\vec d_1.
\eeq

In summary, when we take $\vec d_1$ as fixed, our theory gives a Gaussian 
likelihood function for $\vec d_2$ with mean $\vec\mu$ and 
covariance matrix $\mathbf{M}_c\equiv \mathbf{N}_{22}^{-1}$, both of which depend
on the theory parameters $\vec\theta$. 

It is often convenient to use the block inversion formulae,
\begin{align}
\mathbf{N}_{22}^{-1}&=\mathbf{M}_{22}-\mathbf{M}_{21}\mathbf{M}_{11}^{-1}\mathbf{M}_{21}^T,\\
\mathbf{N}_{22}^{-1}\mathbf{N}_{21}&=-\mathbf{M}_{21}\mathbf{M}_{11}^{-1},
\end{align}
to write
\begin{align}
\vec\mu&=\mathbf{M}_{21}\mathbf{M}_{11}^{-1}\vec d_1,\\
\mathbf{M}_c&=\mathbf{M}_{22}-\mathbf{M}_{21}\mathbf{M}_{11}^{-1}\mathbf{M}_{21}^T.
\end{align}

The full expression for the constrained likelihood is
\beq
p(\vec d_2|\vec d_1,\vec\theta)=
{\exp\left(-{1\over 2}(\vec d_2-\vec\mu(\vec\theta))^T\mathbf{M}_c(\vec\theta)^{-1}
(\vec d_2-\vec\mu(\vec\theta))\right)\over
(2\pi)^{N/2}\mbox{det}^{1/2}\mathbf{M}_c(\vec\theta)},
\eeq
where $N$ is the dimension of $\vec d_2$.

Because we will in general be interested only in this constrained likelihood, we will simplify the notation
by writing $\vec d$ instead of $\vec d_2$ and $\mathbf{M}$ instead of $\mathbf{M}_c$ wherever
there is no risk of ambiguity.

\subsection{Fisher information}

Suppose that we are interested in measuring a single parameter $\theta$, 
such as the modulation level $w$. 
The expected error on $\theta$ is $F^{-1/2}$, where the Fisher information
is 
\beq
F\equiv-\left<(\ln f)''\right>
={1\over 2}\mbox{Tr}(\mathbf{M}^{-1}\mathbf{M}'\mathbf{M}^{-1}\mathbf{M}')+\vec\mu^{\prime T} \mathbf{M}^{-1}\vec\mu',
\label{eq:fisher}
\eeq
and the primes denote derivatives with respect to $\theta$.
In this expression all quantities are to be evaluated at the ``true''
value of $\theta$.

\subsection{Information-maximizing Modes}
\label{sec:modes}

It may be of interest to know what aspects of the new data are most useful in measuring $\theta$. Is it most useful to know large-scale or small-scale information, for instance? Are some parts of the sky more helpful than others? One way to address this sort of question is to suppose that, instead of measuring the entire $N$-dimensional data vector $\vec d$, 
we measure only its projection onto a 
small set of normal modes. To be specific, imagine that we measure
$\delta_j\equiv \vec v_j\cdot\vec d$ for some small set of mode
vectors $\vec v_1,\vec v_2,\ldots$. We can then ask which modes maximize the information in the resulting data set.

For a Gaussian random process whose mean is zero (or more generally, whose mean is independent of $\theta$), the answer to this question is the Karhunen-Lo\`eve transform \cite{karhunen}, which has a long history in cosmology \cite{bondkl,bunnscottwhite,bunnsugiyama,vogeley,vogeleyszalay}. The best modes are the ``signal-to-noise eigenmodes'' with largest eigenvalues. For the constrained data we are considering, the situation is more complicated, as the mean of the distribution depends on the parameter $\theta$.

If there are $K$ mode vectors $\vec v_j$, arranged in the columns of an $N\times K$ matrix $\mathbf{V}$, then the Fisher information in the
data set $\vec\delta$ is
\beq
F_V = {1\over 2}\mathrm{Tr}(\mathbf{M}_V^{-1}\mathbf{M}'_V\mathbf{M}_V^{-1}\mathbf{M}'_V)+
\vec\mu^{\prime T}\mathbf{V}\mathbf{M}_V^{-1}\mathbf{V}^T\vec\mu',
\eeq
where
\beq
\mathbf{M}_V=\mathbf{V}^T\mathbf{M}\mathbf{V},
\qquad
\mathbf{M}'_V=\mathbf{V}^T\mathbf{M}'\mathbf{V}.
\eeq
The information depends only on the subspace spanned by the vectors; i.e., it is invariant under any invertible transformation $\mathbf{V}\to \mathbf{V}\mathbf{A}$. We can therefore without loss of generality choose the vectors to be orthonormal with respect to the inner product 
\beq
\langle \vec x,\vec y\rangle = \vec x^T \mathbf{M}\vec y.
\label{eq:ip}
\eeq
With this choice, $\mathbf{M}_V$ is the identity matrix, and 
\beq
F_V={1\over 2}\mathrm{Tr}((\mathbf{V}^T\mathbf{M}'\mathbf{V})^2)+|\mathbf{V}^T\vec\mu'|^2,
\label{eq:fv}
\eeq

Consider first the case of a single mode ($K=1$), for which
\beq
F_V = {1\over 2}(\vec v^T\mathbf{M}'\vec v)^2 + (\vec v\cdot\vec\mu')^2,
\label{eq:fv1}
\eeq
subject to the constraint $\vec v^T\mathbf{M}\vec v=1$. (When considering the case $K=1$, we omit the subscript on $\vec v_1$.)
We can solve the problem of maximizing $F_V$ by a variety of standard numerical methods, but if one of the two terms in this expression is much larger than the other, then an approximate solution is easily found. The first term satisfies
\beq
{1\over 2}(\vec v^T\mathbf{M}'\vec v)^2 \le {1\over 2}\lambda_{\rm max}^2,
\label{eq:ineq1}
\eeq
where $\lambda_{\rm max}$ is the largest eigenvalue in the generalized eigenvalue problem $\mathbf{M}'\vec u = \lambda\mathbf{M}\vec u$, with
equality when $\vec v$ is the corresponding eigenvector.
The second term satisfies
\beq
(\vec v\cdot\vec\mu')^2 \le (\vec\mu^{\prime T}\mathbf{M}^{-1}\vec\mu')^2
\label{eq:ineq2}
\eeq
with equality when 
\beq
\vec v = {\mathbf{M}^{-1}\vec\mu'\over\sqrt{\vec\mu^{\prime T}
\mathbf{M}^{-1}\vec\mu'}}.
\label{eq:bestmode}
\eeq
 If one of the two expressions on the right side of these inequalities is much larger than the other, then a good approximation to the information-maximizing mode is the mode that saturates that inequality. 

As we will describe in detail in the next section, for all of the cases we consider, the second term is much larger than the first one, and the information-maximizing mode is therefore well-approximated by
equation (\ref{eq:bestmode}).
Moreover, this mode often contains a significant fraction of the total information.
This mode fully captures all of the information that is contained in the way the mean of the probability distribution varies as the parameter $\theta$ is changed. All of the remaining modes will contain information associated only with variations in the covariance matrix.

Having chosen the first mode, we can then seek a second mode that supplies the most additional information. To be specific, let $\vec v_1$ be given by equation (\ref{eq:bestmode}), and let $\vec v_2$ be orthonormal to $\vec v_1$ according to the inner product (\ref{eq:ip}). Then equation (\ref{eq:fv}) can be rewritten
\beq
F_V=F_{\vec v_1}+
{1\over 2}(\vec v_2^T\mathbf{M}'_\perp
\vec v_2)^2+ (\vec v_2\cdot\vec w)^2.
\eeq
Here $F_{\vec v_1}$ is the information contained in mode $\vec v_1$ alone.
The matrix
$\mathbf{M}'_\perp$ is the projection of $\mathbf{M}'$ onto the subspace orthogonal to $\vec v_1$, and $\vec w=\mathbf{M}'\vec v_1$.
Choosing the optimal $\vec v_2$ therefore involves a maximization precisely analogous to that required to find $\vec v_1$ [equation (\ref{eq:fv1})]. This time, however,
as we will see in the next section, the two contributions are comparable for the cases we consider, so neither simple approximate vector is a good solution.  

By an argument analogous to that which led to inequalities (\ref{eq:ineq1}) and (\ref{eq:ineq2}), the new information is bounded by 
\beq
F_{\vec v_2} \le {1\over 2}\lambda_{\perp\,\rm max}^2+(\vec w^T \mathbf{M}_\perp^{-1}\vec w)^2,
\label{eq:ineqv2}
\eeq
where $\lambda_{\perp\,\rm max}$ is the maximum eigenvalue for the generalized eigenvalue problem $\mathbf{M}'_\perp\vec u=\lambda\mathbf{M}_\perp\vec u$.
As we will see, this quantity is small in the cases we will consider, so the second-best mode is of little interest in comparison to the first.

\subsection{Application to CMB polarization}
We will take the previously-measured data $\vec d_1\equiv \vec t$ to be 
CMB temperature data, measured over a masked sky. 
The $j$th measurement can be written
\beq
t_j=\sum_{l,m}a_{lm}^T Y_{lm}(\hat{\bf r}_j)+n_j^T,
\eeq
where $a_{lm}^T$ are the spherical harmonic coefficients,
$\hat{\bf r}_j$ is the location of the $j$th pixel, and $n_j^T$ is the noise.
(For the low-resolution maps we will consider, noise is quite small.
Its primary effect is to regularize the inversion of the covariance
matrix.) 
We write this compactly as
\beq
\vec t = \mathbf{Y}\vec a+\vec n^T.
\eeq
Here the vector $\vec a$ contains the real and imaginary parts of the spherical harmonic coefficients $a_{lm}^T$, and the matrix $\mathbf{Y}$ contains the real and imaginary parts of the corresponding spherical harmonics evaluated at the pixel locations.

The covariance matrix is
\beq
\mathbf{M}_{11}\equiv\langle \vec t\vec t^T\rangle
= \mathbf{Y}\mathbf{C}^{TT}\mathbf{Y}^T+\mathbf{N}_T,
\eeq
where $\mathbf{N}_T$ is the noise covariance matrix and $\mathbf{C}^{TT}$ is the
covariance matrix of the $a_{lm}^T$ coefficients.

The polarization data $\vec d_2$ will consist of polarization measurements,
which can be written
\beq
\vec d_2 = \mathbf{Z}\vec e+\vec n^P.
\eeq
Here $\vec e$ contains the
real and imaginary parts of the E-mode polarization coefficients $a_{lm}^E$,
and $\vec n^P$ is the noise. 
(We neglect the contribution of B modes.) The vector $\vec d_2$ contains the Stokes parameters $Q,U$ for each pixel. The matrix $\mathbf{Z}$ contains the real and imaginary parts of the contributions of the spin-2 spherical harmonics to each $Q$ and $U$ value.

The remaining blocks of the covariance matrix are
\begin{align}
\mathbf{M}_{22}&=\mathbf{Z}\mathbf{C}^{EE}\mathbf{Z}^T+\mathbf{N}_P,\\
\mathbf{M}_{21}&=\mathbf{Z}\mathbf{C}^{ET}\mathbf{Y}^T.
\end{align}
The matrices $\mathbf{C}^{EE}$ and $\mathbf{C}^{ET}$ characterize the covariances of
the $E$ coefficients and the $ET$ cross-covariance, and $\mathbf{N}_P$ is
the noise covariance matrix.

In the standard, statistically-isotropic model, $\mathbf{C}^{TT}, \mathbf{C}^{EE}$,
and $\mathbf{C}^{ET}$ are diagonal matrices containing the three power spectra.
When isotropy is broken, these matrices acquire off-diagonal elements,
which are computed according to the detailed recipe in ref.\ \cite{dvorkin}.
In a coordinate system aligned with the direction of isotropy breaking,
the off-diagonal elements are nonzero only when the two $m$ values are
equal and the $l$'s differ by 1.

\section{Results}
\label{sec:results}

We have performed computations for future polarization data sets, constrained by the existing Planck temperature maps \cite{planck15overview}. To be specific, we used the Planck COMMANDER CMB map with HEALPix \cite{healpix} $N_{\rm side}=256$, downgraded to $N_{\rm side}=32$ and smoothed with a Gaussian beam of width $\sigma_{\rm beam}=2^\circ$. We keep all pixels within the HFI Galactic emission mask, retaining 80\% of the pixels.

We imagine future polarization data with the same smoothing and $N_{\rm side}$,
with signal-to-noise per pixel of 3.
We consider five different sky coverage scenarios:
\begin{itemize}
\item An all-sky map.
\item A map with sky coverage $f_{\rm sky}=0.8$, with a straight Galactic latitude cut, in which all pixels whose Galactic latitude satisfies $|b|>\sin^{-1}(1-f_{\rm sky})$.
\item A map with $f_{\rm sky}=0.6$, with a similar Galactic latitude cut.
\item Two maps with $f_{\rm sky}=0.3$, consisting of spherical caps centered on the North and South Galactic poles. (These two maps combined cover the same area as the $f_{\rm sky}=0.6$ map.)
\end{itemize}

For the broken-isotropy hypothesis, we consider five values for the cutoff wavenumber, namely $k_{\rm max}c/H_0=10,20,30,40,50$. We hold the modulation direction fixed at Galactic coordinates $(l,b)=(226^\circ,-17^\circ)$. All computations are performed after rotating the maps to a coordinate system with this direction at the pole, so that the covariances among the $a_{lm}^{(T,E)}$ coefficients are as simple as possible.

\begin{figure}
\centerline{
\includegraphics[width=3in]{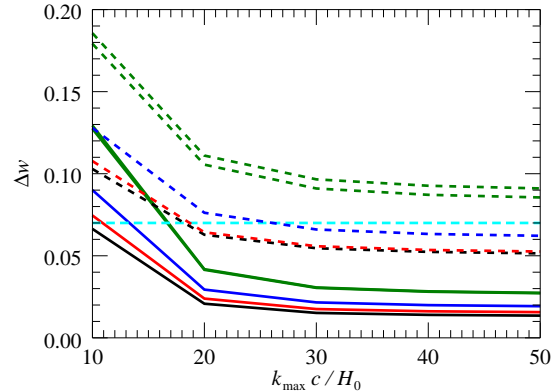}
}
\caption{
The error forecast $\Delta w$. From bottom to top, the solid curves correspond to polarization data sets with $f_{\rm sky}=1$ (black), 0.8 (red), 0.6 (blue), 0.3 (green). 
For $f_{\rm sky}=0.3$, two virtually identical curves are shown, corresponding to the northern and southern caps.
The dashed curves show the error forecasts for a hypothetical experiment in which only the single ``best'' mode of the polarization data is measured.
The horizontal long-dashed line shows the value preferred by the existing temperature data.}
\label{fig:dw}
\end{figure}

The solid curves in Figure \ref{fig:dw} shows the projected error $\Delta w\equiv F^{-1/2}$, where $F$ is the Fisher information. The long-dashed line is the value $w=0.07$, which is roughly the best-fit value from the temperature data. For $k_{\rm max}\gtrsim 30H_0/c$, a strong detection is possible even with relatively low sky coverage.

The dashed curves in the figure show the projected error in the hypothetical scenario where only the information-maximizing mode $\vec v_1$, defined in 
equation (\ref{eq:bestmode}),
is measured. Although this single mode is never enough to provide a definitive measurement, it contains a significant fraction of the total Fisher information, ranging from approximatly 48\% when $k_{\rm max}=10H_0/c$ to 9\% when $k_{\rm max}=50H_0/c$.

As noted in Section \ref{sec:modes}, $\vec v_1$ is in fact an approximation to the information-maximizing mode.
The quality of the approximation is determined by the ratio of the two bounds in inequalities (\ref{eq:ineq1}) and (\ref{eq:ineq2}). For the 
models plotted in the figure, this ratio always exceeds 30, which implies that the information contained in mode $\vec v_1$ is within 3\% of the maximum possible.

The information contained in the second-best mode is bounded by the
inequality (\ref{eq:ineqv2}). In almost all of the cases plotted, the
ratio of this bound to the information contained in the first mode is
less than 8\%, indicating that far more information is contained in
the first mode than in any other individual mode. The only exceptions
occur when $k_{\rm max}=10H_0/c$ and $f_{\rm sky}=0.3$, in which case the total information is quite low. In all cases considered, the two terms in (\ref{eq:ineqv2}) are comparable, differing by no more than a factor of 3, so neither simple approximation would work well for finding the second-best mode. Since this mode is known to contain little information, we do not pursue its calculation further.

\begin{figure*}
\centerline{
\includegraphics[width=1.2in,angle=90]{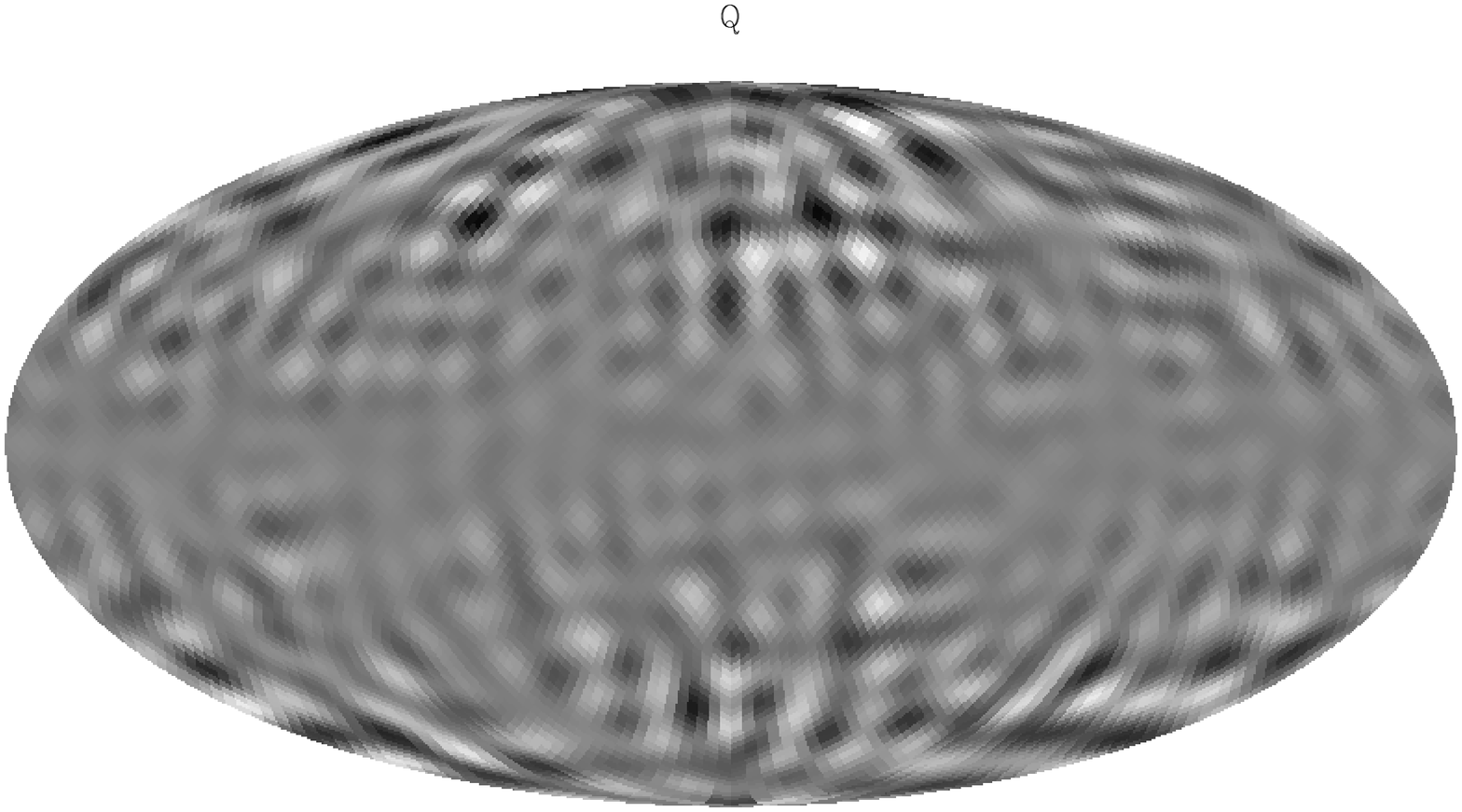}
\hfil
\includegraphics[width=1.2in,angle=90]{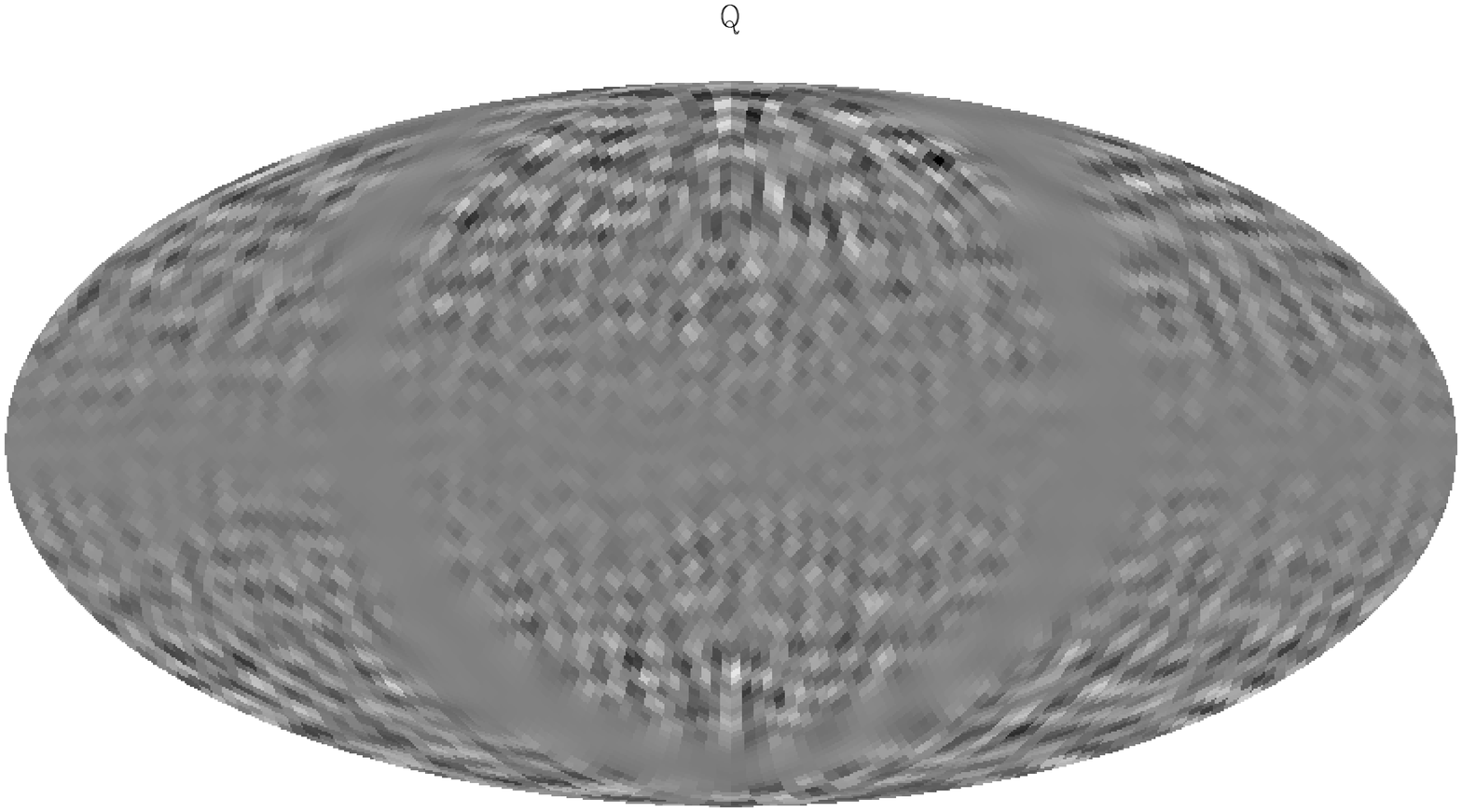}
\hfil
\includegraphics[width=1.2in,angle=90]{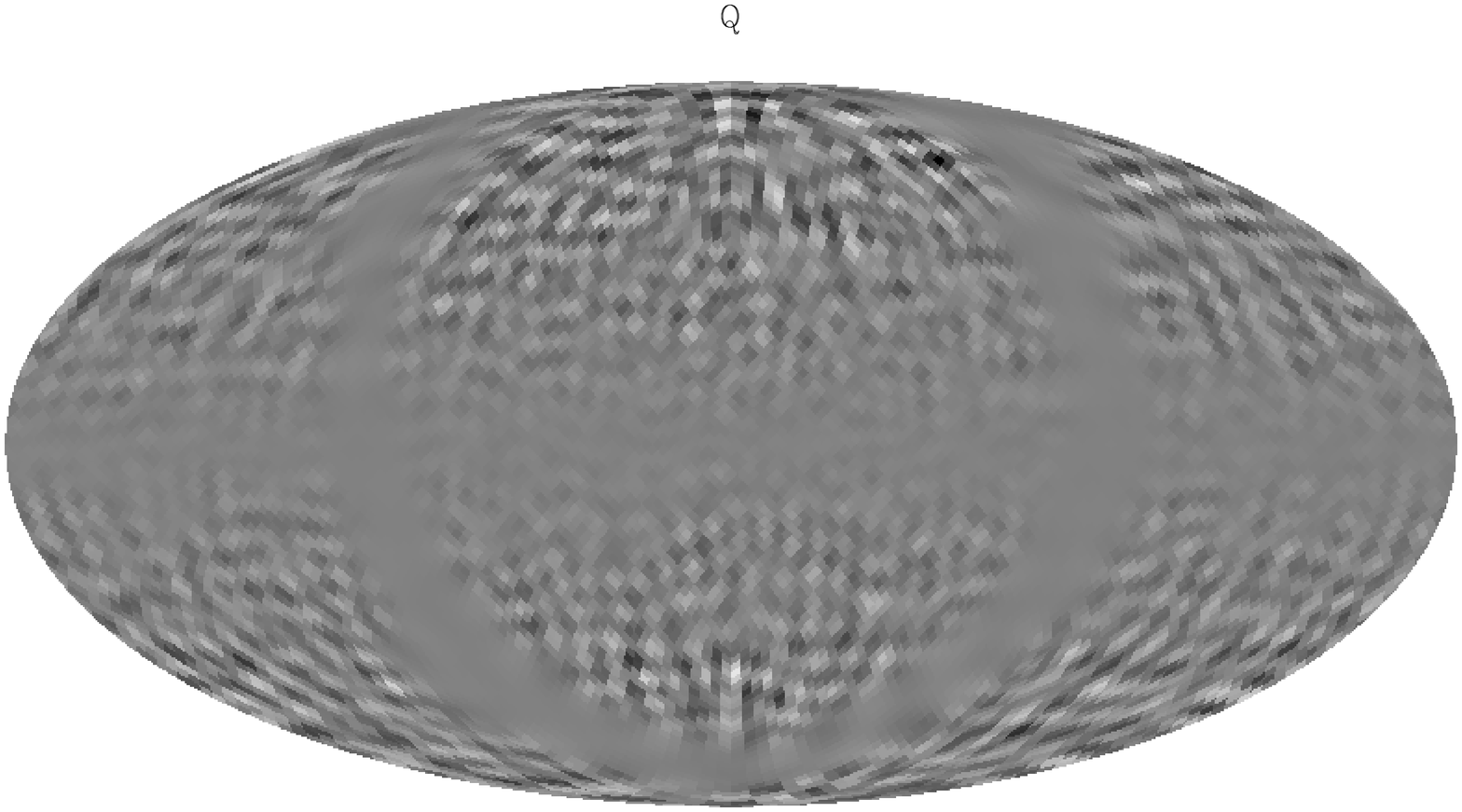}
}
\centerline{
\includegraphics[width=1.2in,angle=90]{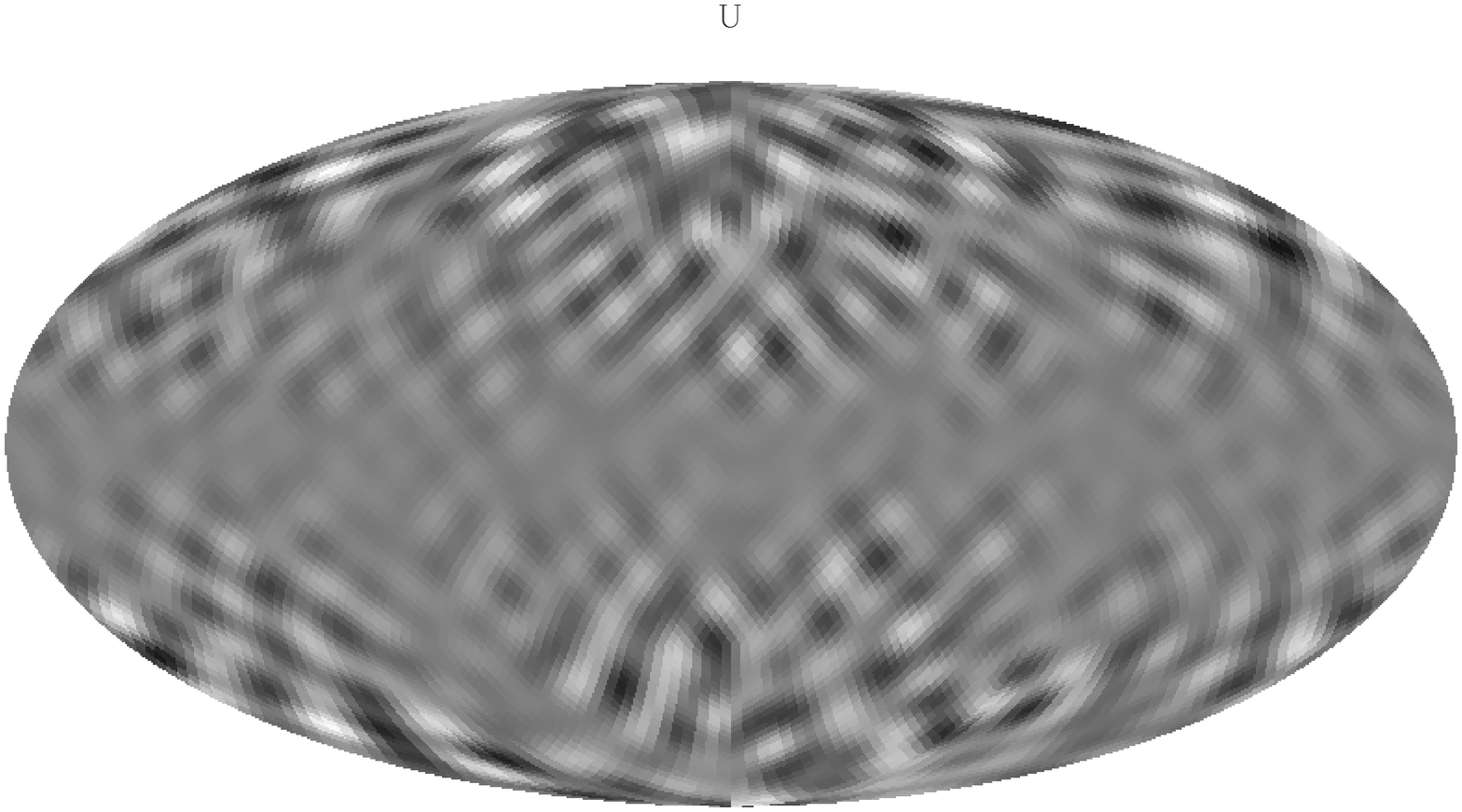}
\hfil
\includegraphics[width=1.2in,angle=90]{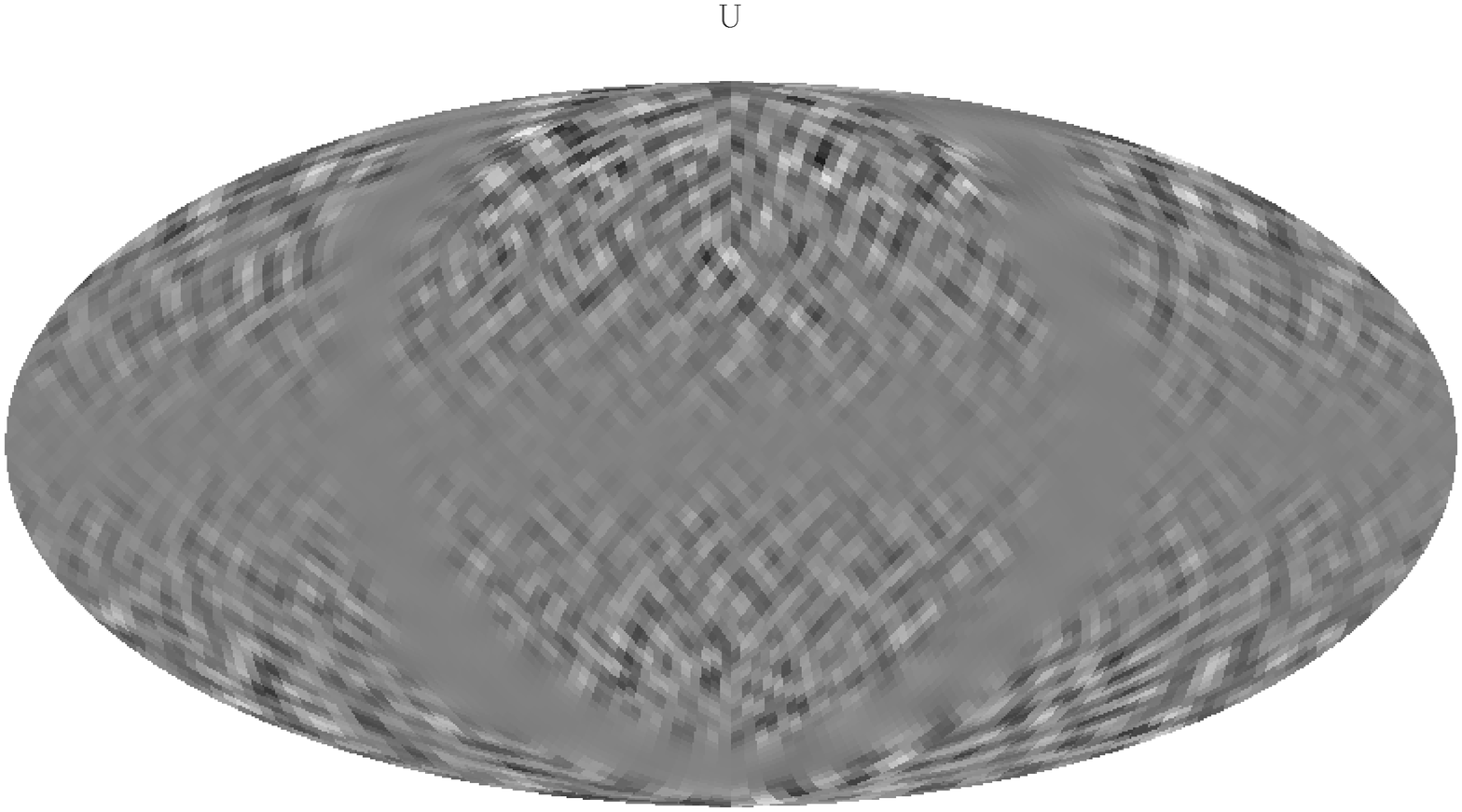}
\hfil
\includegraphics[width=1.2in,angle=90]{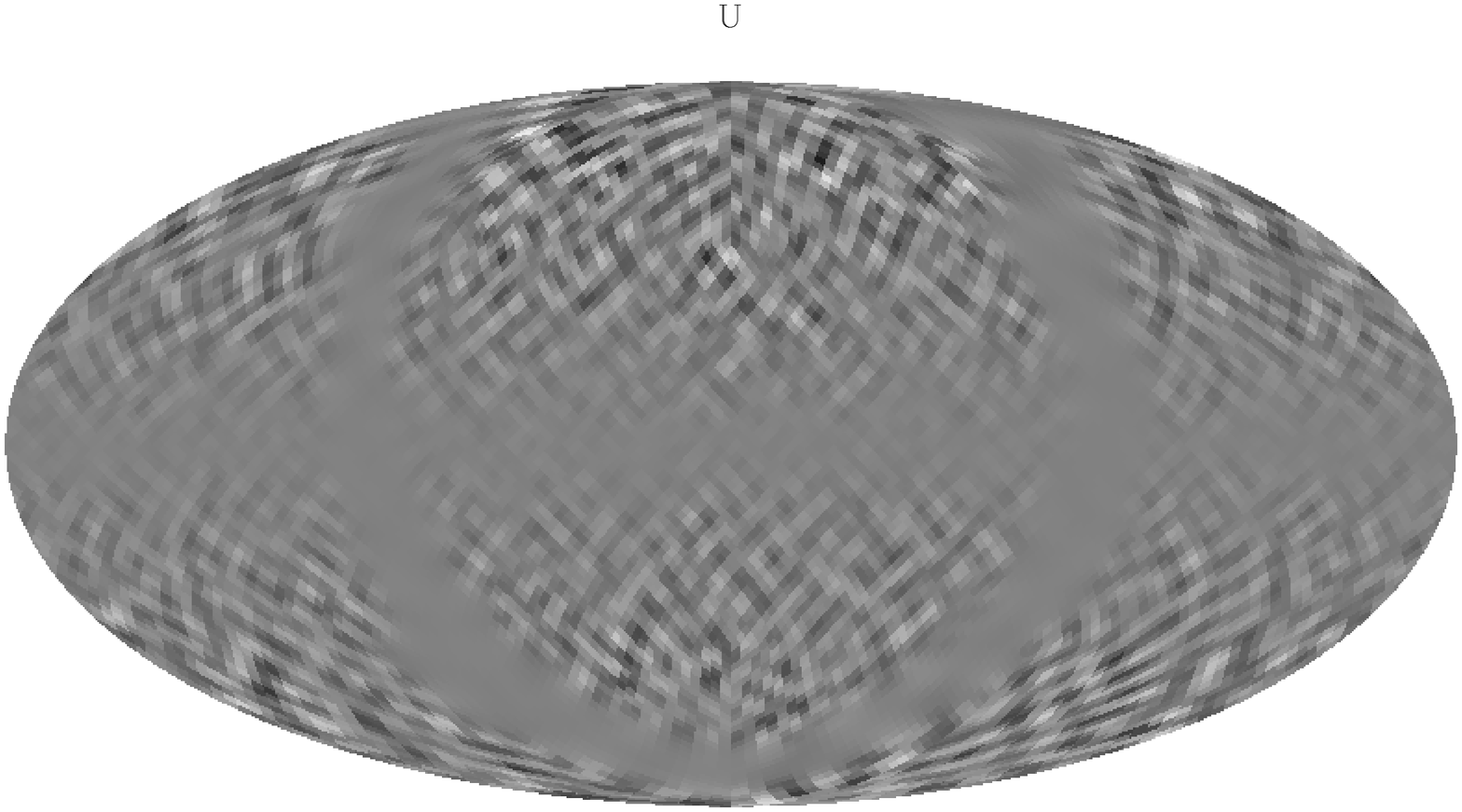}
}
\caption{The single mode that best constrains $w$, for an all-sky polarization data set. The top panel shows Stokes $Q$, and the bottom is Stokes $U$. 
These modes are for an all-sky polarization data set, constrained by the existing temperature data with 80\% sky coverage as described in the text. 
The maps are oriented so that the modulation direction is at the top. The
two bands where the mode is nearly zero are the region of zero modulation (horizontal) and the vicinity of the Galactic plane.
The mode has little power near the Galactic plane because this region is unconstrained by temperature data.
From left to right, $k_{\rm max}=10H_0/c,30H_0/c,50H_0/c$.}
\label{fig:modeskmax}
\end{figure*}

\begin{figure*}
\centerline{
\includegraphics[width=1.2in,angle=90]{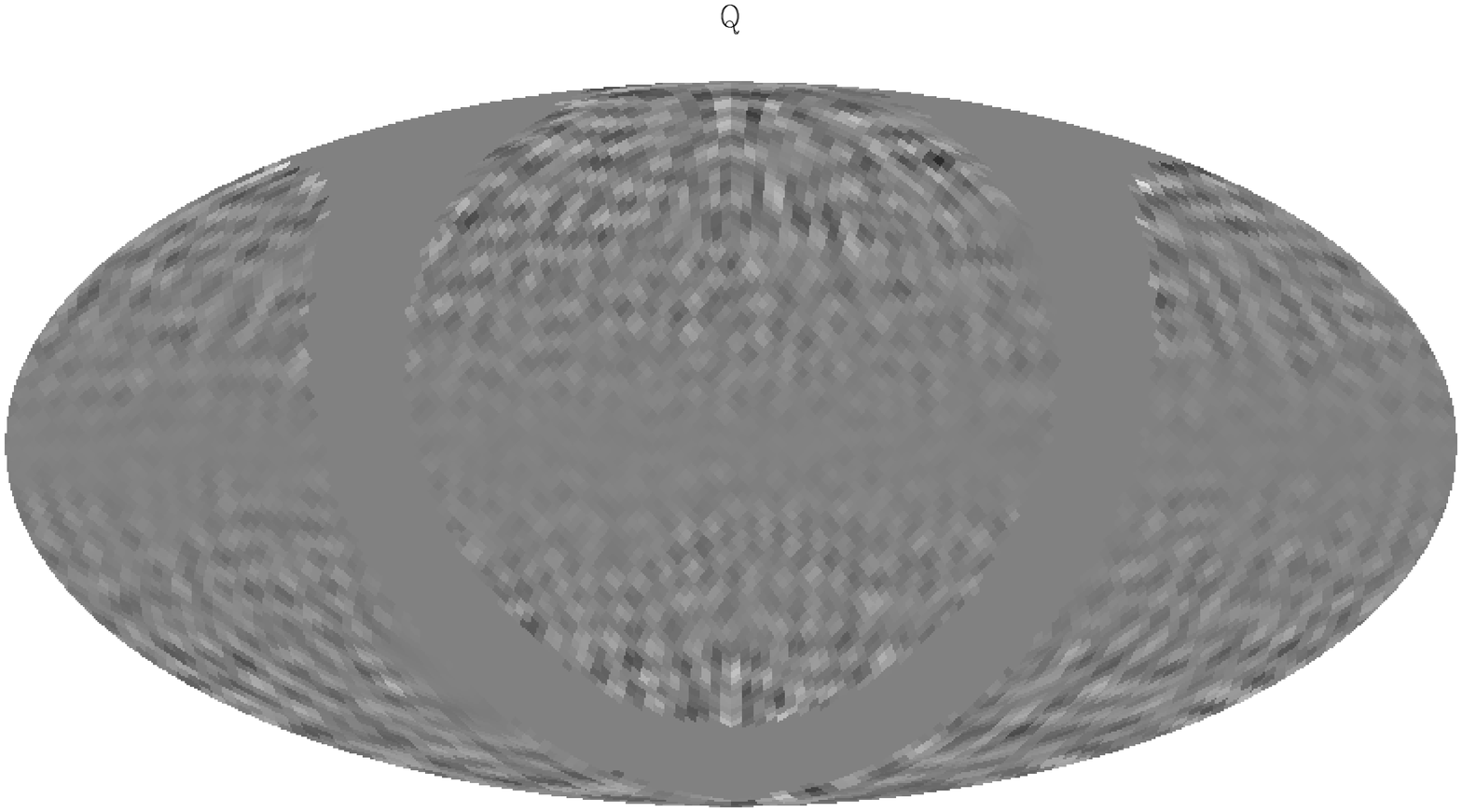}
\hfil
\includegraphics[width=1.2in,angle=90]{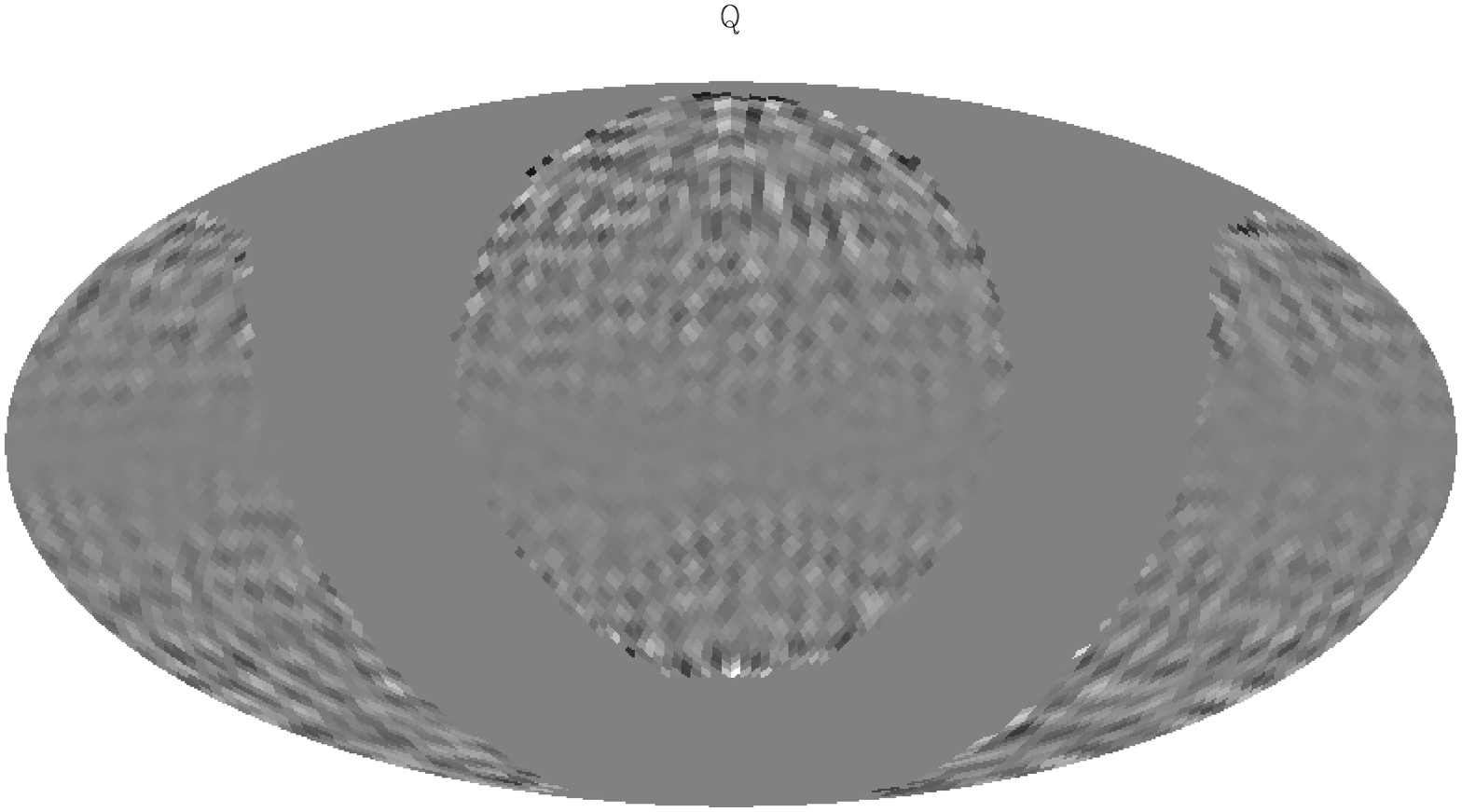}
\hfil
\includegraphics[width=1.2in,angle=90]{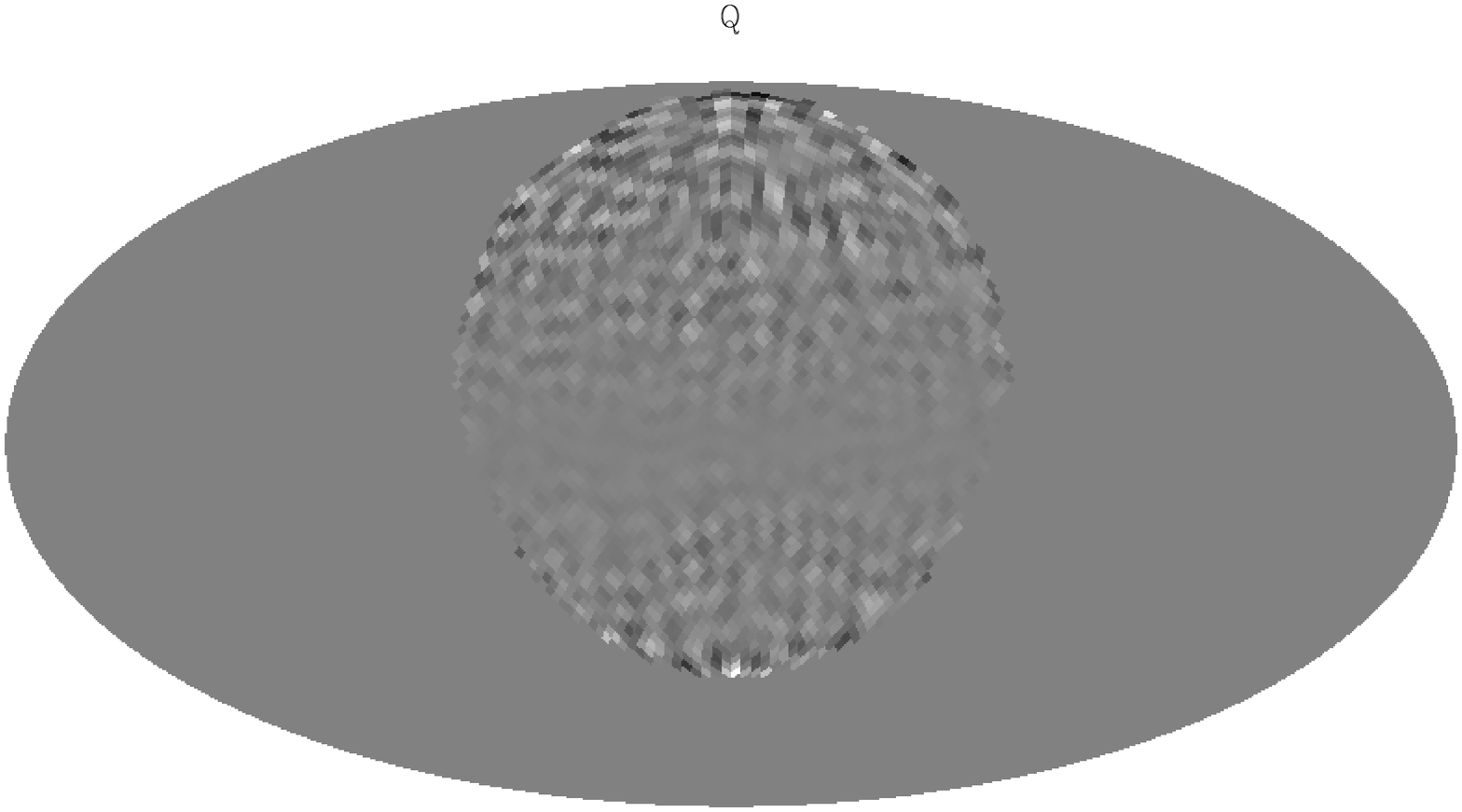}
}
\centerline{
\includegraphics[width=1.2in,angle=90]{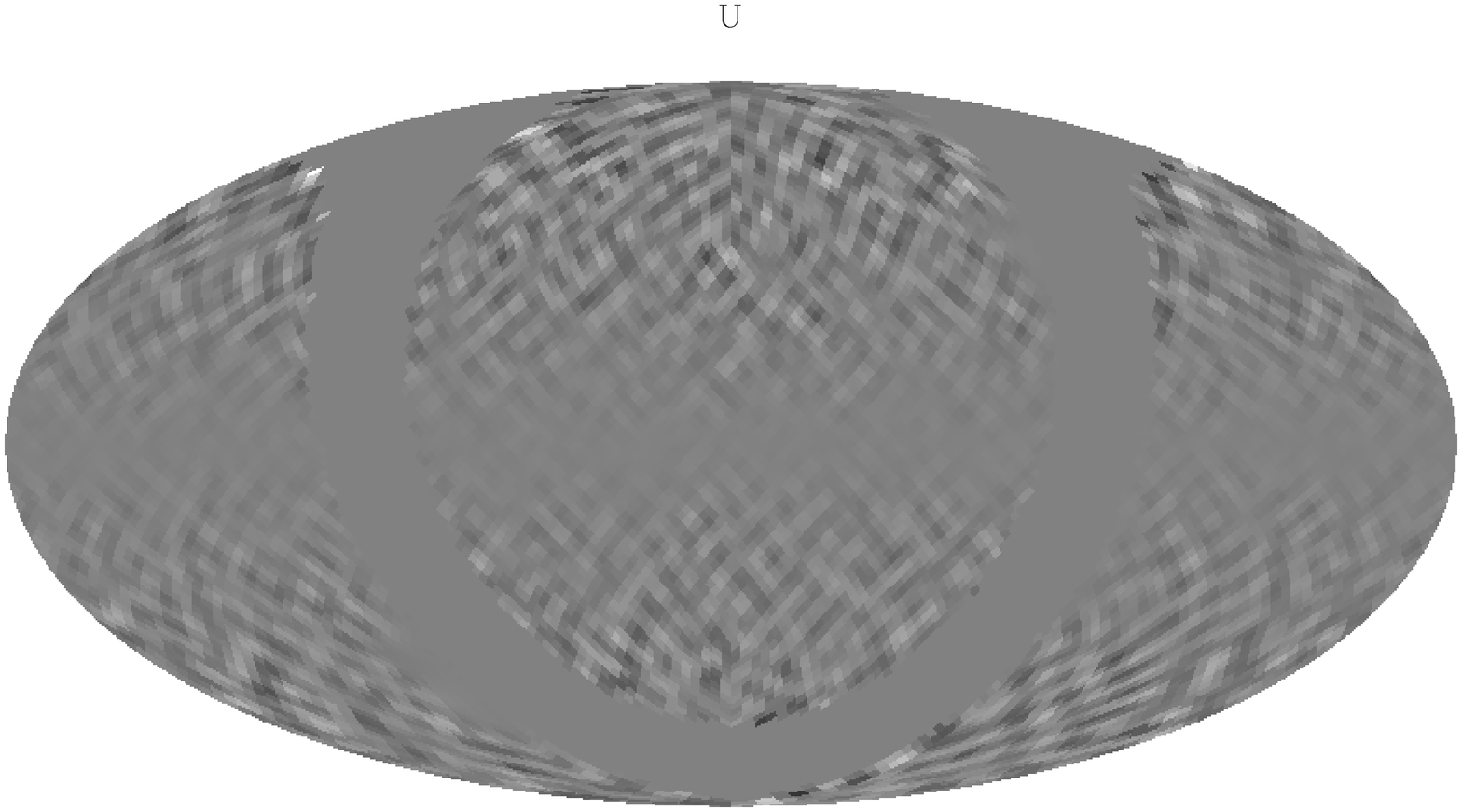}
\hfil
\includegraphics[width=1.2in,angle=90]{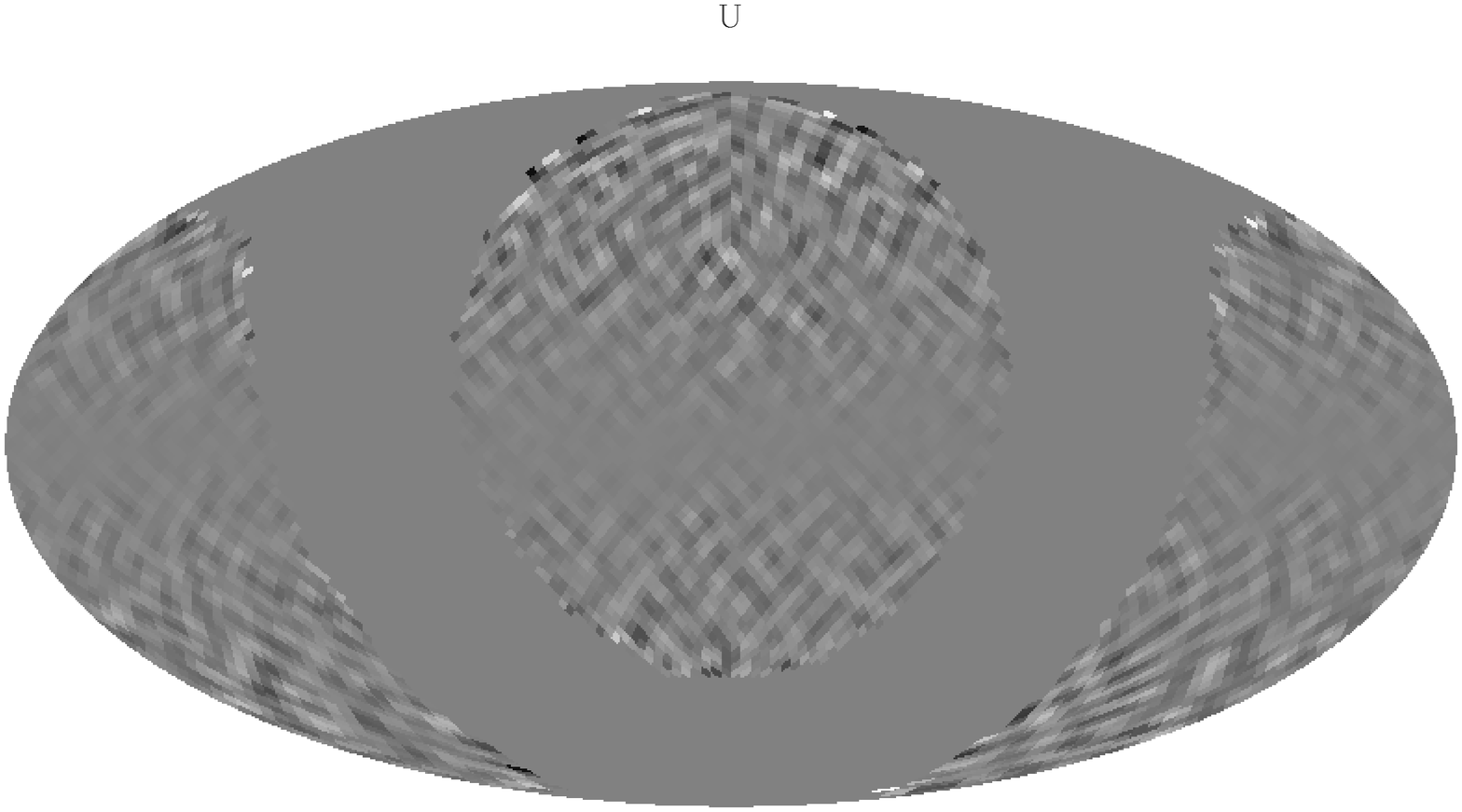}
\hfil
\includegraphics[width=1.2in,angle=90]{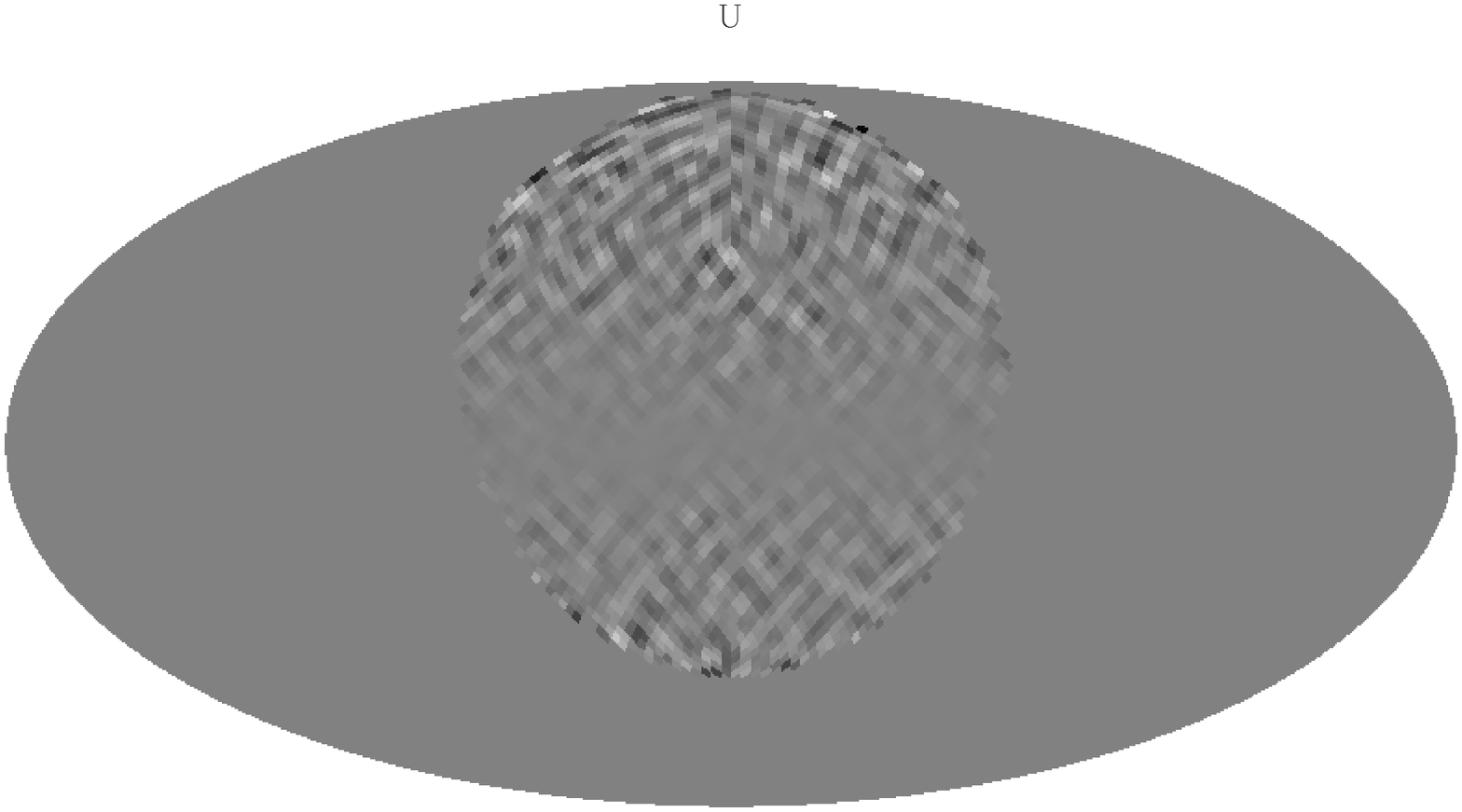}
}
\caption{The best $w$-constraining mode for polarization data sets with sky coverage $f_{\rm sky}=0.8,0.6,0.3$. For $f_{\rm sky}=0.3$, the data are assumed to cover a spherical cap centered on the Galactic north pole; in the other cases, a cut in $|b|$ is used. The maps are oriented as in the previous figure.}
\label{fig:modesfsky}
\end{figure*}

The information-maximizing modes themselves are shown in Figures \ref{fig:modeskmax} and \ref{fig:modesfsky}. Note that these maps are oriented with the modulation direction, rather than the Galactic north pole, at the top. Because these modes are measuring primarily the correlation with the existing temperature data, they have little power in the Galactic plane. Unsurprisingly, they also have little power in the plane perpendicular to the modulation direction, where the modulation is zero.

\section{Discussion}

We have presented Fisher matrix calculations and resulting error forecasts for a future large-scale CMB polarization data set, to assess the degree to which such a data set can shed light on the puzzling large-scale power modulation in the CMB temperature anisotropy. The calculations are based on the probability distribution of the polarization data, conditioned on the already-measured temperature data.

If the observed hemispherical modulation in CMB anisotropy power is not a fluke, a detection of it in CMB polarization should be possible. In the model we have considered, a roughly  3$\sigma$ detection is possible with a data set that covers at least $\sim 60\%$ of the sky, as long as the modulation affects modes with wavenumbers up to $\sim 30H_0/c$. Because these calculations are for the polarization data conditioned on the existing temperature data, this error is associated with the \textit{additional} information available in polarization, on top of what we have already seen in temperature.

The future of CMB polarization measurement on large scales is uncertain, but plans are underway for a ground-based initiative known as CMB-S4 \cite{s4}.
This experiment would be based in the Atacama desert and would survey
a large fraction of the southern sky. The experimenters are also considering including a second telescope in the northern hemisphere to increase the sky coverage. With only southern sky coverage, this experiment would be roughly comparable to our $f_{\rm sky}=0.3$ calculations. With the addition of a northern instrument, the sky coverage would be closer to our $f_{\rm sky}=0.6$.
The calculations for larger sky fractions would correspond roughly
to a hypothetical satellite mission, which would be in the more
distant future. These comparisons are of course extremely rough, as we have not tailored our calculations to match any particular experiment design in detail.

Although the calculations presented herein are based on a simple phenomenological model, we may expect qualitatively similar results from any model in which the observed temperature power modulation is caused by modulation of the large-scale density perturbation modes. 

We have presented a formalism for identifying the spatial modes in the data that best constrain the theory.
A significant fraction (9\% or more) of the information in such a data set is contained in the single ``information-maximizing'' mode that optimally captures the correlation between the known temperature data and the polarization. The rest of the information is contained in the covariances of the polarization measurements (e.g., the predicted mean-square amplitudes of the various modes), although no single mode in this category contains an amount of information comparable to the information-maximizing mode. Because the correlation with the anomalous temperature data is the source of much of the available information, it is necessary to perform constrained calculations as we have done in order to get reliable forecasts.

\section*{Acknowledgments}

This work was supported by NSF awards 0922748 and 1410133. 

\bibliography{polpredict}

\begin{thebibliography}{51}%
\makeatletter
\providecommand \@ifxundefined [1]{%
 \@ifx{#1\undefined}
}%
\providecommand \@ifnum [1]{%
 \ifnum #1\expandafter \@firstoftwo
 \else \expandafter \@secondoftwo
 \fi
}%
\providecommand \@ifx [1]{%
 \ifx #1\expandafter \@firstoftwo
 \else \expandafter \@secondoftwo
 \fi
}%
\providecommand \natexlab [1]{#1}%
\providecommand \enquote  [1]{``#1''}%
\providecommand \bibnamefont  [1]{#1}%
\providecommand \bibfnamefont [1]{#1}%
\providecommand \citenamefont [1]{#1}%
\providecommand \href@noop [0]{\@secondoftwo}%
\providecommand \href [0]{\begingroup \@sanitize@url \@href}%
\providecommand \@href[1]{\@@startlink{#1}\@@href}%
\providecommand \@@href[1]{\endgroup#1\@@endlink}%
\providecommand \@sanitize@url [0]{\catcode `\\12\catcode `\$12\catcode
  `\&12\catcode `\#12\catcode `\^12\catcode `\_12\catcode `\%12\relax}%
\providecommand \@@startlink[1]{}%
\providecommand \@@endlink[0]{}%
\providecommand \url  [0]{\begingroup\@sanitize@url \@url }%
\providecommand \@url [1]{\endgroup\@href {#1}{\urlprefix }}%
\providecommand \urlprefix  [0]{URL }%
\providecommand \Eprint [0]{\href }%
\providecommand \doibase [0]{http://dx.doi.org/}%
\providecommand \selectlanguage [0]{\@gobble}%
\providecommand \bibinfo  [0]{\@secondoftwo}%
\providecommand \bibfield  [0]{\@secondoftwo}%
\providecommand \translation [1]{[#1]}%
\providecommand \BibitemOpen [0]{}%
\providecommand \bibitemStop [0]{}%
\providecommand \bibitemNoStop [0]{.\EOS\space}%
\providecommand \EOS [0]{\spacefactor3000\relax}%
\providecommand \BibitemShut  [1]{\csname bibitem#1\endcsname}%
\let\auto@bib@innerbib\@empty
\bibitem [{\citenamefont {{Planck Collaboration}}\ \emph
  {et~al.}(2015{\natexlab{a}})\citenamefont {{Planck Collaboration}},
  \citenamefont {{Adam}}, \citenamefont {{Ade}}, \citenamefont {{Aghanim}},
  \citenamefont {{Akrami}}, \citenamefont {{Alves}}, \citenamefont {{Arnaud}},
  \citenamefont {{Arroja}}, \citenamefont {{Aumont}}, \citenamefont
  {{Baccigalupi}},\ and\ \citenamefont {et~al.}}]{planck15overview}%
  \BibitemOpen
  \bibfield  {author} {\bibinfo {author} {\bibnamefont {{Planck
  Collaboration}}}, \bibinfo {author} {\bibfnamefont {R.}~\bibnamefont
  {{Adam}}}, \bibinfo {author} {\bibfnamefont {P.~A.~R.}\ \bibnamefont
  {{Ade}}}, \bibinfo {author} {\bibfnamefont {N.}~\bibnamefont {{Aghanim}}},
  \bibinfo {author} {\bibfnamefont {Y.}~\bibnamefont {{Akrami}}}, \bibinfo
  {author} {\bibfnamefont {M.~I.~R.}\ \bibnamefont {{Alves}}}, \bibinfo
  {author} {\bibfnamefont {M.}~\bibnamefont {{Arnaud}}}, \bibinfo {author}
  {\bibfnamefont {F.}~\bibnamefont {{Arroja}}}, \bibinfo {author}
  {\bibfnamefont {J.}~\bibnamefont {{Aumont}}}, \bibinfo {author}
  {\bibfnamefont {C.}~\bibnamefont {{Baccigalupi}}}, \ and\ \bibinfo {author}
  {\bibnamefont {et~al.}},\ }\bibfield  {title} {\enquote {\bibinfo {title}
  {{Planck 2015 results. I. Overview of products and scientific results}},}\
  }\href@noop {} {\bibfield  {journal} {\bibinfo  {journal} {ArXiv e-prints}\ }
  (\bibinfo {year} {2015}{\natexlab{a}})},\ \Eprint
  {http://arxiv.org/abs/1502.01582} {arXiv:1502.01582} \BibitemShut {NoStop}%
\bibitem [{\citenamefont {Bennett}\ \emph {et~al.}(2013)\citenamefont
  {Bennett}, \citenamefont {Larson}, \citenamefont {Weiland}, \citenamefont
  {Jarosik}, \citenamefont {Hinshaw}, \citenamefont {Odegard}, \citenamefont
  {Smith}, \citenamefont {Hill}, \citenamefont {Gold}, \citenamefont {Halpern},
  \citenamefont {Komatsu}, \citenamefont {Nolta}, \citenamefont {Page},
  \citenamefont {Spergel}, \citenamefont {Wollack}, \citenamefont {Dunkley},
  \citenamefont {Kogut}, \citenamefont {Limon}, \citenamefont {Meyer},
  \citenamefont {Tucker},\ and\ \citenamefont {Wright}}]{wmap9yrresults}%
  \BibitemOpen
  \bibfield  {author} {\bibinfo {author} {\bibfnamefont {C.~L.}\ \bibnamefont
  {Bennett}}, \bibinfo {author} {\bibfnamefont {D.}~\bibnamefont {Larson}},
  \bibinfo {author} {\bibfnamefont {J.~L.}\ \bibnamefont {Weiland}}, \bibinfo
  {author} {\bibfnamefont {N.}~\bibnamefont {Jarosik}}, \bibinfo {author}
  {\bibfnamefont {G.}~\bibnamefont {Hinshaw}}, \bibinfo {author} {\bibfnamefont
  {N.}~\bibnamefont {Odegard}}, \bibinfo {author} {\bibfnamefont {K.~M.}\
  \bibnamefont {Smith}}, \bibinfo {author} {\bibfnamefont {R.~S.}\ \bibnamefont
  {Hill}}, \bibinfo {author} {\bibfnamefont {B.}~\bibnamefont {Gold}}, \bibinfo
  {author} {\bibfnamefont {M.}~\bibnamefont {Halpern}}, \bibinfo {author}
  {\bibfnamefont {E.}~\bibnamefont {Komatsu}}, \bibinfo {author} {\bibfnamefont
  {M.~R.}\ \bibnamefont {Nolta}}, \bibinfo {author} {\bibfnamefont
  {L.}~\bibnamefont {Page}}, \bibinfo {author} {\bibfnamefont {D.~N.}\
  \bibnamefont {Spergel}}, \bibinfo {author} {\bibfnamefont {E.}~\bibnamefont
  {Wollack}}, \bibinfo {author} {\bibfnamefont {J.}~\bibnamefont {Dunkley}},
  \bibinfo {author} {\bibfnamefont {A.}~\bibnamefont {Kogut}}, \bibinfo
  {author} {\bibfnamefont {M.}~\bibnamefont {Limon}}, \bibinfo {author}
  {\bibfnamefont {S.~S.}\ \bibnamefont {Meyer}}, \bibinfo {author}
  {\bibfnamefont {G.~S.}\ \bibnamefont {Tucker}}, \ and\ \bibinfo {author}
  {\bibfnamefont {E.~L.}\ \bibnamefont {Wright}},\ }\bibfield  {title}
  {\enquote {\bibinfo {title} {{Nine-year Wilkinson Microwave Anisotropy Probe
  (WMAP) Observations: Final Maps and Results}},}\ }\href
  {http://stacks.iop.org/0067-0049/208/i=2/a=20} {\bibfield  {journal}
  {\bibinfo  {journal} {\apjs}\ }\textbf {\bibinfo {volume} {208}},\ \bibinfo
  {pages} {20} (\bibinfo {year} {2013})}\BibitemShut {NoStop}%
\bibitem [{\citenamefont {{Bucher}}(2015)}]{bucherreview}%
  \BibitemOpen
  \bibfield  {author} {\bibinfo {author} {\bibfnamefont {M.}~\bibnamefont
  {{Bucher}}},\ }\bibfield  {title} {\enquote {\bibinfo {title} {{Physics of
  the cosmic microwave background anisotropy}},}\ }\href {\doibase
  10.1142/S0218271815300049} {\bibfield  {journal} {\bibinfo  {journal} {Int.\
  J.\ Mod.\ Phys.\ D}\ }\textbf {\bibinfo {volume} {24}},\ \bibinfo {eid}
  {1530004-303} (\bibinfo {year} {2015})},\ \Eprint
  {http://arxiv.org/abs/1501.04288} {arXiv:1501.04288} \BibitemShut {NoStop}%
\bibitem [{\citenamefont {{Bennett}}\ \emph {et~al.}(2011)\citenamefont
  {{Bennett}}, \citenamefont {{Hill}}, \citenamefont {{Hinshaw}}, \citenamefont
  {{Larson}}, \citenamefont {{Smith}}, \citenamefont {{Dunkley}}, \citenamefont
  {{Gold}}, \citenamefont {{Halpern}}, \citenamefont {{Jarosik}}, \citenamefont
  {{Kogut}}, \citenamefont {{Komatsu}}, \citenamefont {{Limon}}, \citenamefont
  {{Meyer}}, \citenamefont {{Nolta}}, \citenamefont {{Odegard}}, \citenamefont
  {{Page}}, \citenamefont {{Spergel}}, \citenamefont {{Tucker}}, \citenamefont
  {{Weiland}}, \citenamefont {{Wollack}},\ and\ \citenamefont
  {{Wright}}}]{wmapanomalies}%
  \BibitemOpen
  \bibfield  {author} {\bibinfo {author} {\bibfnamefont {C.~L.}\ \bibnamefont
  {{Bennett}}}, \bibinfo {author} {\bibfnamefont {R.~S.}\ \bibnamefont
  {{Hill}}}, \bibinfo {author} {\bibfnamefont {G.}~\bibnamefont {{Hinshaw}}},
  \bibinfo {author} {\bibfnamefont {D.}~\bibnamefont {{Larson}}}, \bibinfo
  {author} {\bibfnamefont {K.~M.}\ \bibnamefont {{Smith}}}, \bibinfo {author}
  {\bibfnamefont {J.}~\bibnamefont {{Dunkley}}}, \bibinfo {author}
  {\bibfnamefont {B.}~\bibnamefont {{Gold}}}, \bibinfo {author} {\bibfnamefont
  {M.}~\bibnamefont {{Halpern}}}, \bibinfo {author} {\bibfnamefont
  {N.}~\bibnamefont {{Jarosik}}}, \bibinfo {author} {\bibfnamefont
  {A.}~\bibnamefont {{Kogut}}}, \bibinfo {author} {\bibfnamefont
  {E.}~\bibnamefont {{Komatsu}}}, \bibinfo {author} {\bibfnamefont
  {M.}~\bibnamefont {{Limon}}}, \bibinfo {author} {\bibfnamefont {S.~S.}\
  \bibnamefont {{Meyer}}}, \bibinfo {author} {\bibfnamefont {M.~R.}\
  \bibnamefont {{Nolta}}}, \bibinfo {author} {\bibfnamefont {N.}~\bibnamefont
  {{Odegard}}}, \bibinfo {author} {\bibfnamefont {L.}~\bibnamefont {{Page}}},
  \bibinfo {author} {\bibfnamefont {D.~N.}\ \bibnamefont {{Spergel}}}, \bibinfo
  {author} {\bibfnamefont {G.~S.}\ \bibnamefont {{Tucker}}}, \bibinfo {author}
  {\bibfnamefont {J.~L.}\ \bibnamefont {{Weiland}}}, \bibinfo {author}
  {\bibfnamefont {E.}~\bibnamefont {{Wollack}}}, \ and\ \bibinfo {author}
  {\bibfnamefont {E.~L.}\ \bibnamefont {{Wright}}},\ }\bibfield  {title}
  {\enquote {\bibinfo {title} {{Seven-year Wilkinson Microwave Anisotropy Probe
  (WMAP) Observations: Are There Cosmic Microwave Background Anomalies?}}}\
  }\href {\doibase 10.1088/0067-0049/192/2/17} {\bibfield  {journal} {\bibinfo
  {journal} {\apjs}\ }\textbf {\bibinfo {volume} {192}},\ \bibinfo {eid} {17}
  (\bibinfo {year} {2011})},\ \Eprint {http://arxiv.org/abs/1001.4758}
  {arXiv:1001.4758 [astro-ph.CO]} \BibitemShut {NoStop}%
\bibitem [{\citenamefont {{Planck Collaboration}}\ \emph
  {et~al.}(2014)\citenamefont {{Planck Collaboration}}, \citenamefont {{Ade}},
  \citenamefont {{Aghanim}}, \citenamefont {{Armitage-Caplan}}, \citenamefont
  {{Arnaud}}, \citenamefont {{Ashdown}}, \citenamefont {{Atrio-Barandela}},
  \citenamefont {{Aumont}}, \citenamefont {{Baccigalupi}}, \citenamefont
  {{Banday}},\ and\ \citenamefont {et~al.}}]{planck13isotropy}%
  \BibitemOpen
  \bibfield  {author} {\bibinfo {author} {\bibnamefont {{Planck
  Collaboration}}}, \bibinfo {author} {\bibfnamefont {P.~A.~R.}\ \bibnamefont
  {{Ade}}}, \bibinfo {author} {\bibfnamefont {N.}~\bibnamefont {{Aghanim}}},
  \bibinfo {author} {\bibfnamefont {C.}~\bibnamefont {{Armitage-Caplan}}},
  \bibinfo {author} {\bibfnamefont {M.}~\bibnamefont {{Arnaud}}}, \bibinfo
  {author} {\bibfnamefont {M.}~\bibnamefont {{Ashdown}}}, \bibinfo {author}
  {\bibfnamefont {F.}~\bibnamefont {{Atrio-Barandela}}}, \bibinfo {author}
  {\bibfnamefont {J.}~\bibnamefont {{Aumont}}}, \bibinfo {author}
  {\bibfnamefont {C.}~\bibnamefont {{Baccigalupi}}}, \bibinfo {author}
  {\bibfnamefont {A.~J.}\ \bibnamefont {{Banday}}}, \ and\ \bibinfo {author}
  {\bibnamefont {et~al.}},\ }\bibfield  {title} {\enquote {\bibinfo {title}
  {{Planck 2013 results. XXIII. Isotropy and statistics of the CMB}},}\ }\href
  {\doibase 10.1051/0004-6361/201321534} {\bibfield  {journal} {\bibinfo
  {journal} {\aap}\ }\textbf {\bibinfo {volume} {571}},\ \bibinfo {eid} {A23}
  (\bibinfo {year} {2014})},\ \Eprint {http://arxiv.org/abs/1303.5083}
  {arXiv:1303.5083} \BibitemShut {NoStop}%
\bibitem [{\citenamefont {{Planck Collaboration}}\ \emph
  {et~al.}(2015{\natexlab{b}})\citenamefont {{Planck Collaboration}},
  \citenamefont {{Ade}}, \citenamefont {{Aghanim}}, \citenamefont {{Akrami}},
  \citenamefont {{Aluri}}, \citenamefont {{Arnaud}}, \citenamefont {{Ashdown}},
  \citenamefont {{Aumont}}, \citenamefont {{Baccigalupi}}, \citenamefont
  {{Banday}},\ and\ \citenamefont {et~al.}}]{planck15isotropy}%
  \BibitemOpen
  \bibfield  {author} {\bibinfo {author} {\bibnamefont {{Planck
  Collaboration}}}, \bibinfo {author} {\bibfnamefont {P.~A.~R.}\ \bibnamefont
  {{Ade}}}, \bibinfo {author} {\bibfnamefont {N.}~\bibnamefont {{Aghanim}}},
  \bibinfo {author} {\bibfnamefont {Y.}~\bibnamefont {{Akrami}}}, \bibinfo
  {author} {\bibfnamefont {P.~K.}\ \bibnamefont {{Aluri}}}, \bibinfo {author}
  {\bibfnamefont {M.}~\bibnamefont {{Arnaud}}}, \bibinfo {author}
  {\bibfnamefont {M.}~\bibnamefont {{Ashdown}}}, \bibinfo {author}
  {\bibfnamefont {J.}~\bibnamefont {{Aumont}}}, \bibinfo {author}
  {\bibfnamefont {C.}~\bibnamefont {{Baccigalupi}}}, \bibinfo {author}
  {\bibfnamefont {A.~J.}\ \bibnamefont {{Banday}}}, \ and\ \bibinfo {author}
  {\bibnamefont {et~al.}},\ }\bibfield  {title} {\enquote {\bibinfo {title}
  {{Planck 2015 results. XVI. Isotropy and statistics of the CMB}},}\
  }\href@noop {} {\bibfield  {journal} {\bibinfo  {journal} {ArXiv e-prints}\ }
  (\bibinfo {year} {2015}{\natexlab{b}})},\ \Eprint
  {http://arxiv.org/abs/1506.07135} {arXiv:1506.07135} \BibitemShut {NoStop}%
\bibitem [{\citenamefont {{Schwarz}}\ \emph {et~al.}(2004)\citenamefont
  {{Schwarz}}, \citenamefont {{Starkman}}, \citenamefont {{Huterer}},\ and\
  \citenamefont {{Copi}}}]{schwarz04alignment}%
  \BibitemOpen
  \bibfield  {author} {\bibinfo {author} {\bibfnamefont {D.~J.}\ \bibnamefont
  {{Schwarz}}}, \bibinfo {author} {\bibfnamefont {G.~D.}\ \bibnamefont
  {{Starkman}}}, \bibinfo {author} {\bibfnamefont {D.}~\bibnamefont
  {{Huterer}}}, \ and\ \bibinfo {author} {\bibfnamefont {C.~J.}\ \bibnamefont
  {{Copi}}},\ }\bibfield  {title} {\enquote {\bibinfo {title} {{Is the
  Low-{$\ell$} Microwave Background Cosmic?}}}\ }\href {\doibase
  10.1103/PhysRevLett.93.221301} {\bibfield  {journal} {\bibinfo  {journal}
  {\prl}\ }\textbf {\bibinfo {volume} {93}},\ \bibinfo {eid} {221301} (\bibinfo
  {year} {2004})},\ \Eprint {http://arxiv.org/abs/astro-ph/0403353}
  {astro-ph/0403353} \BibitemShut {NoStop}%
\bibitem [{\citenamefont {{Copi}}\ \emph {et~al.}(2006)\citenamefont {{Copi}},
  \citenamefont {{Huterer}}, \citenamefont {{Schwarz}},\ and\ \citenamefont
  {{Starkman}}}]{copi06}%
  \BibitemOpen
  \bibfield  {author} {\bibinfo {author} {\bibfnamefont {C.~J.}\ \bibnamefont
  {{Copi}}}, \bibinfo {author} {\bibfnamefont {D.}~\bibnamefont {{Huterer}}},
  \bibinfo {author} {\bibfnamefont {D.~J.}\ \bibnamefont {{Schwarz}}}, \ and\
  \bibinfo {author} {\bibfnamefont {G.~D.}\ \bibnamefont {{Starkman}}},\
  }\bibfield  {title} {\enquote {\bibinfo {title} {{On the large-angle
  anomalies of the microwave sky}},}\ }\href {\doibase
  10.1111/j.1365-2966.2005.09980.x} {\bibfield  {journal} {\bibinfo  {journal}
  {\mnras}\ }\textbf {\bibinfo {volume} {367}},\ \bibinfo {pages} {79--102}
  (\bibinfo {year} {2006})},\ \Eprint {http://arxiv.org/abs/astro-ph/0508047}
  {astro-ph/0508047} \BibitemShut {NoStop}%
\bibitem [{\citenamefont {{Schwarz}}\ \emph {et~al.}(2015)\citenamefont
  {{Schwarz}}, \citenamefont {{Copi}}, \citenamefont {{Huterer}},\ and\
  \citenamefont {{Starkman}}}]{schwarz15}%
  \BibitemOpen
  \bibfield  {author} {\bibinfo {author} {\bibfnamefont {D.~J.}\ \bibnamefont
  {{Schwarz}}}, \bibinfo {author} {\bibfnamefont {C.~J.}\ \bibnamefont
  {{Copi}}}, \bibinfo {author} {\bibfnamefont {D.}~\bibnamefont {{Huterer}}}, \
  and\ \bibinfo {author} {\bibfnamefont {G.~D.}\ \bibnamefont {{Starkman}}},\
  }\bibfield  {title} {\enquote {\bibinfo {title} {{CMB Anomalies after
  Planck}},}\ }\href@noop {} {\bibfield  {journal} {\bibinfo  {journal} {ArXiv
  e-prints}\ } (\bibinfo {year} {2015})},\ \Eprint
  {http://arxiv.org/abs/1510.07929} {arXiv:1510.07929} \BibitemShut {NoStop}%
\bibitem [{\citenamefont {{de Oliveira-Costa}}\ \emph
  {et~al.}(2004)\citenamefont {{de Oliveira-Costa}}, \citenamefont {{Tegmark}},
  \citenamefont {{Zaldarriaga}},\ and\ \citenamefont {{Hamilton}}}]{dOTZH}%
  \BibitemOpen
  \bibfield  {author} {\bibinfo {author} {\bibfnamefont {A.}~\bibnamefont {{de
  Oliveira-Costa}}}, \bibinfo {author} {\bibfnamefont {M.}~\bibnamefont
  {{Tegmark}}}, \bibinfo {author} {\bibfnamefont {M.}~\bibnamefont
  {{Zaldarriaga}}}, \ and\ \bibinfo {author} {\bibfnamefont {A.}~\bibnamefont
  {{Hamilton}}},\ }\bibfield  {title} {\enquote {\bibinfo {title}
  {{Significance of the largest scale CMB fluctuations in WMAP}},}\ }\href
  {\doibase 10.1103/PhysRevD.69.063516} {\bibfield  {journal} {\bibinfo
  {journal} {\prd}\ }\textbf {\bibinfo {volume} {69}},\ \bibinfo {eid} {063516}
  (\bibinfo {year} {2004})},\ \Eprint {http://arxiv.org/abs/astro-ph/0307282}
  {astro-ph/0307282} \BibitemShut {NoStop}%
\bibitem [{\citenamefont {{Copi}}\ \emph {et~al.}(2004)\citenamefont {{Copi}},
  \citenamefont {{Huterer}},\ and\ \citenamefont {{Starkman}}}]{copimv}%
  \BibitemOpen
  \bibfield  {author} {\bibinfo {author} {\bibfnamefont {C.~J.}\ \bibnamefont
  {{Copi}}}, \bibinfo {author} {\bibfnamefont {D.}~\bibnamefont {{Huterer}}}, \
  and\ \bibinfo {author} {\bibfnamefont {G.~D.}\ \bibnamefont {{Starkman}}},\
  }\bibfield  {title} {\enquote {\bibinfo {title} {{Multipole vectors: A new
  representation of the CMB sky and evidence for statistical anisotropy or
  non-Gaussianity at $2{\le}l{\le}8$}},}\ }\href {\doibase
  10.1103/PhysRevD.70.043515} {\bibfield  {journal} {\bibinfo  {journal}
  {\prd}\ }\textbf {\bibinfo {volume} {70}},\ \bibinfo {eid} {043515} (\bibinfo
  {year} {2004})},\ \Eprint {http://arxiv.org/abs/astro-ph/0310511}
  {astro-ph/0310511} \BibitemShut {NoStop}%
\bibitem [{\citenamefont {{Land}}\ and\ \citenamefont
  {{Magueijo}}(2005)}]{landmagueijo}%
  \BibitemOpen
  \bibfield  {author} {\bibinfo {author} {\bibfnamefont {K.}~\bibnamefont
  {{Land}}}\ and\ \bibinfo {author} {\bibfnamefont {J.}~\bibnamefont
  {{Magueijo}}},\ }\bibfield  {title} {\enquote {\bibinfo {title} {{The
  multipole vectors of the Wilkinson Microwave Anisotropy Probe, and their
  frames and invariants}},}\ }\href {\doibase 10.1111/j.1365-2966.2005.09310.x}
  {\bibfield  {journal} {\bibinfo  {journal} {\mnras}\ }\textbf {\bibinfo
  {volume} {362}},\ \bibinfo {pages} {838--846} (\bibinfo {year} {2005})},\
  \Eprint {http://arxiv.org/abs/astro-ph/0502574} {astro-ph/0502574}
  \BibitemShut {NoStop}%
\bibitem [{\citenamefont {{Copi}}\ \emph {et~al.}(2015)\citenamefont {{Copi}},
  \citenamefont {{Huterer}}, \citenamefont {{Schwarz}},\ and\ \citenamefont
  {{Starkman}}}]{copi15alignment}%
  \BibitemOpen
  \bibfield  {author} {\bibinfo {author} {\bibfnamefont {C.~J.}\ \bibnamefont
  {{Copi}}}, \bibinfo {author} {\bibfnamefont {D.}~\bibnamefont {{Huterer}}},
  \bibinfo {author} {\bibfnamefont {D.~J.}\ \bibnamefont {{Schwarz}}}, \ and\
  \bibinfo {author} {\bibfnamefont {G.~D.}\ \bibnamefont {{Starkman}}},\
  }\bibfield  {title} {\enquote {\bibinfo {title} {{Large-scale alignments from
  WMAP and Planck}},}\ }\href {\doibase 10.1093/mnras/stv501} {\bibfield
  {journal} {\bibinfo  {journal} {\mnras}\ }\textbf {\bibinfo {volume} {449}},\
  \bibinfo {pages} {3458--3470} (\bibinfo {year} {2015})},\ \Eprint
  {http://arxiv.org/abs/1311.4562} {arXiv:1311.4562} \BibitemShut {NoStop}%
\bibitem [{\citenamefont {{Eriksen}}\ \emph {et~al.}(2004)\citenamefont
  {{Eriksen}}, \citenamefont {{Hansen}}, \citenamefont {{Banday}},
  \citenamefont {{G{\'o}rski}},\ and\ \citenamefont {{Lilje}}}]{eriksen04}%
  \BibitemOpen
  \bibfield  {author} {\bibinfo {author} {\bibfnamefont {H.~K.}\ \bibnamefont
  {{Eriksen}}}, \bibinfo {author} {\bibfnamefont {F.~K.}\ \bibnamefont
  {{Hansen}}}, \bibinfo {author} {\bibfnamefont {A.~J.}\ \bibnamefont
  {{Banday}}}, \bibinfo {author} {\bibfnamefont {K.~M.}\ \bibnamefont
  {{G{\'o}rski}}}, \ and\ \bibinfo {author} {\bibfnamefont {P.~B.}\
  \bibnamefont {{Lilje}}},\ }\bibfield  {title} {\enquote {\bibinfo {title}
  {{Asymmetries in the Cosmic Microwave Background Anisotropy Field}},}\ }\href
  {\doibase 10.1086/382267} {\bibfield  {journal} {\bibinfo  {journal} {\apj}\
  }\textbf {\bibinfo {volume} {605}},\ \bibinfo {pages} {14--20} (\bibinfo
  {year} {2004})},\ \Eprint {http://arxiv.org/abs/astro-ph/0307507}
  {astro-ph/0307507} \BibitemShut {NoStop}%
\bibitem [{\citenamefont {{Eriksen}}\ \emph {et~al.}(2007)\citenamefont
  {{Eriksen}}, \citenamefont {{Banday}}, \citenamefont {{G{\'o}rski}},
  \citenamefont {{Hansen}},\ and\ \citenamefont {{Lilje}}}]{eriksen07}%
  \BibitemOpen
  \bibfield  {author} {\bibinfo {author} {\bibfnamefont {H.~K.}\ \bibnamefont
  {{Eriksen}}}, \bibinfo {author} {\bibfnamefont {A.~J.}\ \bibnamefont
  {{Banday}}}, \bibinfo {author} {\bibfnamefont {K.~M.}\ \bibnamefont
  {{G{\'o}rski}}}, \bibinfo {author} {\bibfnamefont {F.~K.}\ \bibnamefont
  {{Hansen}}}, \ and\ \bibinfo {author} {\bibfnamefont {P.~B.}\ \bibnamefont
  {{Lilje}}},\ }\bibfield  {title} {\enquote {\bibinfo {title} {{Hemispherical
  Power Asymmetry in the Third-Year Wilkinson Microwave Anisotropy Probe Sky
  Maps}},}\ }\href {\doibase 10.1086/518091} {\bibfield  {journal} {\bibinfo
  {journal} {\apjl}\ }\textbf {\bibinfo {volume} {660}},\ \bibinfo {pages}
  {L81--L84} (\bibinfo {year} {2007})},\ \Eprint
  {http://arxiv.org/abs/astro-ph/0701089} {astro-ph/0701089} \BibitemShut
  {NoStop}%
\bibitem [{\citenamefont {{Gordon}}(2007)}]{gordon07}%
  \BibitemOpen
  \bibfield  {author} {\bibinfo {author} {\bibfnamefont {C.}~\bibnamefont
  {{Gordon}}},\ }\bibfield  {title} {\enquote {\bibinfo {title} {{Broken
  Isotropy from a Linear Modulation of the Primordial Perturbations}},}\ }\href
  {\doibase 10.1086/510511} {\bibfield  {journal} {\bibinfo  {journal} {\apj}\
  }\textbf {\bibinfo {volume} {656}},\ \bibinfo {pages} {636--640} (\bibinfo
  {year} {2007})},\ \Eprint {http://arxiv.org/abs/astro-ph/0607423}
  {astro-ph/0607423} \BibitemShut {NoStop}%
\bibitem [{\citenamefont {{Hansen}}\ \emph {et~al.}(2009)\citenamefont
  {{Hansen}}, \citenamefont {{Banday}}, \citenamefont {{G{\'o}rski}},
  \citenamefont {{Eriksen}},\ and\ \citenamefont {{Lilje}}}]{hansen09}%
  \BibitemOpen
  \bibfield  {author} {\bibinfo {author} {\bibfnamefont {F.~K.}\ \bibnamefont
  {{Hansen}}}, \bibinfo {author} {\bibfnamefont {A.~J.}\ \bibnamefont
  {{Banday}}}, \bibinfo {author} {\bibfnamefont {K.~M.}\ \bibnamefont
  {{G{\'o}rski}}}, \bibinfo {author} {\bibfnamefont {H.~K.}\ \bibnamefont
  {{Eriksen}}}, \ and\ \bibinfo {author} {\bibfnamefont {P.~B.}\ \bibnamefont
  {{Lilje}}},\ }\bibfield  {title} {\enquote {\bibinfo {title} {{Power
  Asymmetry in Cosmic Microwave Background Fluctuations from Full Sky to
  Sub-Degree Scales: Is the Universe Isotropic?}}}\ }\href {\doibase
  10.1088/0004-637X/704/2/1448} {\bibfield  {journal} {\bibinfo  {journal}
  {\apj}\ }\textbf {\bibinfo {volume} {704}},\ \bibinfo {pages} {1448--1458}
  (\bibinfo {year} {2009})},\ \Eprint {http://arxiv.org/abs/0812.3795}
  {arXiv:0812.3795} \BibitemShut {NoStop}%
\bibitem [{\citenamefont {{Hoftuft}}\ \emph {et~al.}(2009)\citenamefont
  {{Hoftuft}}, \citenamefont {{Eriksen}}, \citenamefont {{Banday}},
  \citenamefont {{G{\'o}rski}}, \citenamefont {{Hansen}},\ and\ \citenamefont
  {{Lilje}}}]{hoftuft09}%
  \BibitemOpen
  \bibfield  {author} {\bibinfo {author} {\bibfnamefont {J.}~\bibnamefont
  {{Hoftuft}}}, \bibinfo {author} {\bibfnamefont {H.~K.}\ \bibnamefont
  {{Eriksen}}}, \bibinfo {author} {\bibfnamefont {A.~J.}\ \bibnamefont
  {{Banday}}}, \bibinfo {author} {\bibfnamefont {K.~M.}\ \bibnamefont
  {{G{\'o}rski}}}, \bibinfo {author} {\bibfnamefont {F.~K.}\ \bibnamefont
  {{Hansen}}}, \ and\ \bibinfo {author} {\bibfnamefont {P.~B.}\ \bibnamefont
  {{Lilje}}},\ }\bibfield  {title} {\enquote {\bibinfo {title} {{Increasing
  Evidence for Hemispherical Power Asymmetry in the Five-Year WMAP Data}},}\
  }\href {\doibase 10.1088/0004-637X/699/2/985} {\bibfield  {journal} {\bibinfo
   {journal} {\apj}\ }\textbf {\bibinfo {volume} {699}},\ \bibinfo {pages}
  {985--989} (\bibinfo {year} {2009})},\ \Eprint
  {http://arxiv.org/abs/0903.1229} {arXiv:0903.1229 [astro-ph.CO]} \BibitemShut
  {NoStop}%
\bibitem [{\citenamefont {{Akrami}}\ \emph {et~al.}(2014)\citenamefont
  {{Akrami}}, \citenamefont {{Fantaye}}, \citenamefont {{Shafieloo}},
  \citenamefont {{Eriksen}}, \citenamefont {{Hansen}}, \citenamefont
  {{Banday}},\ and\ \citenamefont {{G{\'o}rski}}}]{akrami14}%
  \BibitemOpen
  \bibfield  {author} {\bibinfo {author} {\bibfnamefont {Y.}~\bibnamefont
  {{Akrami}}}, \bibinfo {author} {\bibfnamefont {Y.}~\bibnamefont {{Fantaye}}},
  \bibinfo {author} {\bibfnamefont {A.}~\bibnamefont {{Shafieloo}}}, \bibinfo
  {author} {\bibfnamefont {H.~K.}\ \bibnamefont {{Eriksen}}}, \bibinfo {author}
  {\bibfnamefont {F.~K.}\ \bibnamefont {{Hansen}}}, \bibinfo {author}
  {\bibfnamefont {A.~J.}\ \bibnamefont {{Banday}}}, \ and\ \bibinfo {author}
  {\bibfnamefont {K.~M.}\ \bibnamefont {{G{\'o}rski}}},\ }\bibfield  {title}
  {\enquote {\bibinfo {title} {{Power Asymmetry in WMAP and Planck Temperature
  Sky Maps as Measured by a Local Variance Estimator}},}\ }\href {\doibase
  10.1088/2041-8205/784/2/L42} {\bibfield  {journal} {\bibinfo  {journal}
  {\apjl}\ }\textbf {\bibinfo {volume} {784}},\ \bibinfo {eid} {L42} (\bibinfo
  {year} {2014})},\ \Eprint {http://arxiv.org/abs/1402.0870} {arXiv:1402.0870}
  \BibitemShut {NoStop}%
\bibitem [{\citenamefont {{Adhikari}}(2015)}]{adhikari15}%
  \BibitemOpen
  \bibfield  {author} {\bibinfo {author} {\bibfnamefont {S.}~\bibnamefont
  {{Adhikari}}},\ }\bibfield  {title} {\enquote {\bibinfo {title} {{Local
  variance asymmetries in Planck temperature anisotropy maps}},}\ }\href
  {\doibase 10.1093/mnras/stu2408} {\bibfield  {journal} {\bibinfo  {journal}
  {\mnras}\ }\textbf {\bibinfo {volume} {446}},\ \bibinfo {pages} {4232--4238}
  (\bibinfo {year} {2015})},\ \Eprint {http://arxiv.org/abs/1408.5396}
  {arXiv:1408.5396} \BibitemShut {NoStop}%
\bibitem [{\citenamefont {{Aiola}}\ \emph {et~al.}(2015)\citenamefont
  {{Aiola}}, \citenamefont {{Wang}}, \citenamefont {{Kosowsky}}, \citenamefont
  {{Kahniashvili}},\ and\ \citenamefont {{Firouzjahi}}}]{aiola15}%
  \BibitemOpen
  \bibfield  {author} {\bibinfo {author} {\bibfnamefont {S.}~\bibnamefont
  {{Aiola}}}, \bibinfo {author} {\bibfnamefont {B.}~\bibnamefont {{Wang}}},
  \bibinfo {author} {\bibfnamefont {A.}~\bibnamefont {{Kosowsky}}}, \bibinfo
  {author} {\bibfnamefont {T.}~\bibnamefont {{Kahniashvili}}}, \ and\ \bibinfo
  {author} {\bibfnamefont {H.}~\bibnamefont {{Firouzjahi}}},\ }\bibfield
  {title} {\enquote {\bibinfo {title} {{Microwave background correlations from
  dipole anisotropy modulation}},}\ }\href {\doibase
  10.1103/PhysRevD.92.063008} {\bibfield  {journal} {\bibinfo  {journal}
  {\prd}\ }\textbf {\bibinfo {volume} {92}},\ \bibinfo {eid} {063008} (\bibinfo
  {year} {2015})},\ \Eprint {http://arxiv.org/abs/1506.04405}
  {arXiv:1506.04405} \BibitemShut {NoStop}%
\bibitem [{\citenamefont {{Efstathiou}}(2003)}]{efstathiou}%
  \BibitemOpen
  \bibfield  {author} {\bibinfo {author} {\bibfnamefont {G.}~\bibnamefont
  {{Efstathiou}}},\ }\bibfield  {title} {\enquote {\bibinfo {title} {{The
  statistical significance of the low cosmic microwave background
  mulitipoles}},}\ }\href {\doibase 10.1046/j.1365-2966.2003.07304.x}
  {\bibfield  {journal} {\bibinfo  {journal} {\mnras}\ }\textbf {\bibinfo
  {volume} {346}},\ \bibinfo {pages} {L26--L30} (\bibinfo {year} {2003})},\
  \Eprint {http://arxiv.org/abs/astro-ph/0306431} {astro-ph/0306431}
  \BibitemShut {NoStop}%
\bibitem [{\citenamefont {{Paci}}\ \emph {et~al.}(2010)\citenamefont {{Paci}},
  \citenamefont {{Gruppuso}}, \citenamefont {{Finelli}}, \citenamefont
  {{Cabella}}, \citenamefont {{de Rosa}}, \citenamefont {{Mandolesi}},\ and\
  \citenamefont {{Natoli}}}]{paci10}%
  \BibitemOpen
  \bibfield  {author} {\bibinfo {author} {\bibfnamefont {F.}~\bibnamefont
  {{Paci}}}, \bibinfo {author} {\bibfnamefont {A.}~\bibnamefont {{Gruppuso}}},
  \bibinfo {author} {\bibfnamefont {F.}~\bibnamefont {{Finelli}}}, \bibinfo
  {author} {\bibfnamefont {P.}~\bibnamefont {{Cabella}}}, \bibinfo {author}
  {\bibfnamefont {A.}~\bibnamefont {{de Rosa}}}, \bibinfo {author}
  {\bibfnamefont {N.}~\bibnamefont {{Mandolesi}}}, \ and\ \bibinfo {author}
  {\bibfnamefont {P.}~\bibnamefont {{Natoli}}},\ }\bibfield  {title} {\enquote
  {\bibinfo {title} {{Power asymmetries in the cosmic microwave background
  temperature and polarization patterns}},}\ }\href {\doibase
  10.1111/j.1365-2966.2010.16905.x} {\bibfield  {journal} {\bibinfo  {journal}
  {\mnras}\ }\textbf {\bibinfo {volume} {407}},\ \bibinfo {pages} {399--404}
  (\bibinfo {year} {2010})},\ \Eprint {http://arxiv.org/abs/1002.4745}
  {arXiv:1002.4745 [astro-ph.CO]} \BibitemShut {NoStop}%
\bibitem [{\citenamefont {{Paci}}\ \emph {et~al.}(2013)\citenamefont {{Paci}},
  \citenamefont {{Gruppuso}}, \citenamefont {{Finelli}}, \citenamefont {{De
  Rosa}}, \citenamefont {{Mandolesi}},\ and\ \citenamefont
  {{Natoli}}}]{paci13}%
  \BibitemOpen
  \bibfield  {author} {\bibinfo {author} {\bibfnamefont {F.}~\bibnamefont
  {{Paci}}}, \bibinfo {author} {\bibfnamefont {A.}~\bibnamefont {{Gruppuso}}},
  \bibinfo {author} {\bibfnamefont {F.}~\bibnamefont {{Finelli}}}, \bibinfo
  {author} {\bibfnamefont {A.}~\bibnamefont {{De Rosa}}}, \bibinfo {author}
  {\bibfnamefont {N.}~\bibnamefont {{Mandolesi}}}, \ and\ \bibinfo {author}
  {\bibfnamefont {P.}~\bibnamefont {{Natoli}}},\ }\bibfield  {title} {\enquote
  {\bibinfo {title} {{Hemispherical power asymmetries in the WMAP 7-year
  low-resolution temperature and polarization maps}},}\ }\href {\doibase
  10.1093/mnras/stt1219} {\bibfield  {journal} {\bibinfo  {journal} {\mnras}\
  }\textbf {\bibinfo {volume} {434}},\ \bibinfo {pages} {3071--3077} (\bibinfo
  {year} {2013})},\ \Eprint {http://arxiv.org/abs/1301.5195} {arXiv:1301.5195}
  \BibitemShut {NoStop}%
\bibitem [{\citenamefont {{Copi}}\ \emph {et~al.}(2013)\citenamefont {{Copi}},
  \citenamefont {{Huterer}}, \citenamefont {{Schwarz}},\ and\ \citenamefont
  {{Starkman}}}]{copi13}%
  \BibitemOpen
  \bibfield  {author} {\bibinfo {author} {\bibfnamefont {C.~J.}\ \bibnamefont
  {{Copi}}}, \bibinfo {author} {\bibfnamefont {D.}~\bibnamefont {{Huterer}}},
  \bibinfo {author} {\bibfnamefont {D.~J.}\ \bibnamefont {{Schwarz}}}, \ and\
  \bibinfo {author} {\bibfnamefont {G.~D.}\ \bibnamefont {{Starkman}}},\
  }\bibfield  {title} {\enquote {\bibinfo {title} {{Large-angle cosmic
  microwave background suppression and polarization predictions}},}\ }\href
  {\doibase 10.1093/mnras/stt1287} {\bibfield  {journal} {\bibinfo  {journal}
  {\mnras}\ }\textbf {\bibinfo {volume} {434}},\ \bibinfo {pages} {3590--3596}
  (\bibinfo {year} {2013})},\ \Eprint {http://arxiv.org/abs/1303.4786}
  {arXiv:1303.4786} \BibitemShut {NoStop}%
\bibitem [{\citenamefont {{Yoho}}\ \emph {et~al.}(2015)\citenamefont {{Yoho}},
  \citenamefont {{Aiola}}, \citenamefont {{Copi}}, \citenamefont {{Kosowsky}},\
  and\ \citenamefont {{Starkman}}}]{yoho}%
  \BibitemOpen
  \bibfield  {author} {\bibinfo {author} {\bibfnamefont {A.}~\bibnamefont
  {{Yoho}}}, \bibinfo {author} {\bibfnamefont {S.}~\bibnamefont {{Aiola}}},
  \bibinfo {author} {\bibfnamefont {C.~J.}\ \bibnamefont {{Copi}}}, \bibinfo
  {author} {\bibfnamefont {A.}~\bibnamefont {{Kosowsky}}}, \ and\ \bibinfo
  {author} {\bibfnamefont {G.~D.}\ \bibnamefont {{Starkman}}},\ }\bibfield
  {title} {\enquote {\bibinfo {title} {{Microwave background polarization as a
  probe of large-angle correlations}},}\ }\href {\doibase
  10.1103/PhysRevD.91.123504} {\bibfield  {journal} {\bibinfo  {journal}
  {\prd}\ }\textbf {\bibinfo {volume} {91}},\ \bibinfo {eid} {123504} (\bibinfo
  {year} {2015})},\ \Eprint {http://arxiv.org/abs/1503.05928}
  {arXiv:1503.05928} \BibitemShut {NoStop}%
\bibitem [{\citenamefont {{O'Dwyer}}\ \emph {et~al.}(2016)\citenamefont
  {{O'Dwyer}}, \citenamefont {{Copi}}, \citenamefont {{Knox}},\ and\
  \citenamefont {{Starkman}}}]{odwyer}%
  \BibitemOpen
  \bibfield  {author} {\bibinfo {author} {\bibfnamefont {M.}~\bibnamefont
  {{O'Dwyer}}}, \bibinfo {author} {\bibfnamefont {C.~J.}\ \bibnamefont
  {{Copi}}}, \bibinfo {author} {\bibfnamefont {L.}~\bibnamefont {{Knox}}}, \
  and\ \bibinfo {author} {\bibfnamefont {G.~D.}\ \bibnamefont {{Starkman}}},\
  }\bibfield  {title} {\enquote {\bibinfo {title} {{CMB-S4 and the
  Hemispherical Variance Anomaly}},}\ }\href@noop {} {\bibfield  {journal}
  {\bibinfo  {journal} {ArXiv e-prints}\ } (\bibinfo {year} {2016})},\ \Eprint
  {http://arxiv.org/abs/1608.02234} {arXiv:1608.02234} \BibitemShut {NoStop}%
\bibitem [{\citenamefont {{Dvorkin}}\ \emph {et~al.}(2008)\citenamefont
  {{Dvorkin}}, \citenamefont {{Peiris}},\ and\ \citenamefont {{Hu}}}]{dvorkin}%
  \BibitemOpen
  \bibfield  {author} {\bibinfo {author} {\bibfnamefont {C.}~\bibnamefont
  {{Dvorkin}}}, \bibinfo {author} {\bibfnamefont {H.~V.}\ \bibnamefont
  {{Peiris}}}, \ and\ \bibinfo {author} {\bibfnamefont {W.}~\bibnamefont
  {{Hu}}},\ }\bibfield  {title} {\enquote {\bibinfo {title} {{Testable
  polarization predictions for models of CMB isotropy anomalies}},}\ }\href
  {\doibase 10.1103/PhysRevD.77.063008} {\bibfield  {journal} {\bibinfo
  {journal} {\prd}\ }\textbf {\bibinfo {volume} {77}},\ \bibinfo {eid} {063008}
  (\bibinfo {year} {2008})},\ \Eprint {http://arxiv.org/abs/0711.2321}
  {arXiv:0711.2321} \BibitemShut {NoStop}%
\bibitem [{\citenamefont {{Erickcek}}\ \emph
  {et~al.}(2008{\natexlab{a}})\citenamefont {{Erickcek}}, \citenamefont
  {{Carroll}},\ and\ \citenamefont {{Kamionkowski}}}]{erickcek08}%
  \BibitemOpen
  \bibfield  {author} {\bibinfo {author} {\bibfnamefont {A.~L.}\ \bibnamefont
  {{Erickcek}}}, \bibinfo {author} {\bibfnamefont {S.~M.}\ \bibnamefont
  {{Carroll}}}, \ and\ \bibinfo {author} {\bibfnamefont {M.}~\bibnamefont
  {{Kamionkowski}}},\ }\bibfield  {title} {\enquote {\bibinfo {title}
  {{Superhorizon perturbations and the cosmic microwave background}},}\ }\href
  {\doibase 10.1103/PhysRevD.78.083012} {\bibfield  {journal} {\bibinfo
  {journal} {\prd}\ }\textbf {\bibinfo {volume} {78}},\ \bibinfo {eid} {083012}
  (\bibinfo {year} {2008}{\natexlab{a}})},\ \Eprint
  {http://arxiv.org/abs/0808.1570} {arXiv:0808.1570} \BibitemShut {NoStop}%
\bibitem [{\citenamefont {{Erickcek}}\ \emph
  {et~al.}(2008{\natexlab{b}})\citenamefont {{Erickcek}}, \citenamefont
  {{Kamionkowski}},\ and\ \citenamefont {{Carroll}}}]{erickcek08b}%
  \BibitemOpen
  \bibfield  {author} {\bibinfo {author} {\bibfnamefont {A.~L.}\ \bibnamefont
  {{Erickcek}}}, \bibinfo {author} {\bibfnamefont {M.}~\bibnamefont
  {{Kamionkowski}}}, \ and\ \bibinfo {author} {\bibfnamefont {S.~M.}\
  \bibnamefont {{Carroll}}},\ }\bibfield  {title} {\enquote {\bibinfo {title}
  {{A hemispherical power asymmetry from inflation}},}\ }\href {\doibase
  10.1103/PhysRevD.78.123520} {\bibfield  {journal} {\bibinfo  {journal}
  {\prd}\ }\textbf {\bibinfo {volume} {78}},\ \bibinfo {eid} {123520} (\bibinfo
  {year} {2008}{\natexlab{b}})},\ \Eprint {http://arxiv.org/abs/0806.0377}
  {arXiv:0806.0377} \BibitemShut {NoStop}%
\bibitem [{\citenamefont {{Erickcek}}\ \emph {et~al.}(2009)\citenamefont
  {{Erickcek}}, \citenamefont {{Hirata}},\ and\ \citenamefont
  {{Kamionkowski}}}]{erickcek09}%
  \BibitemOpen
  \bibfield  {author} {\bibinfo {author} {\bibfnamefont {A.~L.}\ \bibnamefont
  {{Erickcek}}}, \bibinfo {author} {\bibfnamefont {C.~M.}\ \bibnamefont
  {{Hirata}}}, \ and\ \bibinfo {author} {\bibfnamefont {M.}~\bibnamefont
  {{Kamionkowski}}},\ }\bibfield  {title} {\enquote {\bibinfo {title} {{A
  scale-dependent power asymmetry from isocurvature perturbations}},}\ }\href
  {\doibase 10.1103/PhysRevD.80.083507} {\bibfield  {journal} {\bibinfo
  {journal} {\prd}\ }\textbf {\bibinfo {volume} {80}},\ \bibinfo {eid} {083507}
  (\bibinfo {year} {2009})},\ \Eprint {http://arxiv.org/abs/0907.0705}
  {arXiv:0907.0705 [astro-ph.CO]} \BibitemShut {NoStop}%
\bibitem [{\citenamefont {{Moss}}\ \emph {et~al.}(2011)\citenamefont {{Moss}},
  \citenamefont {{Scott}}, \citenamefont {{Zibin}},\ and\ \citenamefont
  {{Battye}}}]{moss}%
  \BibitemOpen
  \bibfield  {author} {\bibinfo {author} {\bibfnamefont {A.}~\bibnamefont
  {{Moss}}}, \bibinfo {author} {\bibfnamefont {D.}~\bibnamefont {{Scott}}},
  \bibinfo {author} {\bibfnamefont {J.~P.}\ \bibnamefont {{Zibin}}}, \ and\
  \bibinfo {author} {\bibfnamefont {R.}~\bibnamefont {{Battye}}},\ }\bibfield
  {title} {\enquote {\bibinfo {title} {{Tilted physics: A cosmologically
  dipole-modulated sky}},}\ }\href {\doibase 10.1103/PhysRevD.84.023014}
  {\bibfield  {journal} {\bibinfo  {journal} {\prd}\ }\textbf {\bibinfo
  {volume} {84}},\ \bibinfo {eid} {023014} (\bibinfo {year} {2011})},\ \Eprint
  {http://arxiv.org/abs/1011.2990} {arXiv:1011.2990 [astro-ph.CO]} \BibitemShut
  {NoStop}%
\bibitem [{\citenamefont {{Zibin}}\ and\ \citenamefont
  {{Contreras}}(2015)}]{zibin}%
  \BibitemOpen
  \bibfield  {author} {\bibinfo {author} {\bibfnamefont {J.~P.}\ \bibnamefont
  {{Zibin}}}\ and\ \bibinfo {author} {\bibfnamefont {D.}~\bibnamefont
  {{Contreras}}},\ }\bibfield  {title} {\enquote {\bibinfo {title} {{Testing
  physical models for dipolar asymmetry: from temperature to k space to
  lensing}},}\ }\href@noop {} {\bibfield  {journal} {\bibinfo  {journal} {ArXiv
  e-prints}\ } (\bibinfo {year} {2015})},\ \Eprint
  {http://arxiv.org/abs/1512.02618} {arXiv:1512.02618} \BibitemShut {NoStop}%
\bibitem [{\citenamefont {{Byrnes}}\ \emph {et~al.}(2016)\citenamefont
  {{Byrnes}}, \citenamefont {{Regan}}, \citenamefont {{Seery}},\ and\
  \citenamefont {{Tarrant}}}]{byrnes}%
  \BibitemOpen
  \bibfield  {author} {\bibinfo {author} {\bibfnamefont {C.~T.}\ \bibnamefont
  {{Byrnes}}}, \bibinfo {author} {\bibfnamefont {D.}~\bibnamefont {{Regan}}},
  \bibinfo {author} {\bibfnamefont {D.}~\bibnamefont {{Seery}}}, \ and\
  \bibinfo {author} {\bibfnamefont {E.~R.~M.}\ \bibnamefont {{Tarrant}}},\
  }\bibfield  {title} {\enquote {\bibinfo {title} {{Implications of the cosmic
  microwave background power asymmetry for the early universe}},}\ }\href
  {\doibase 10.1103/PhysRevD.93.123003} {\bibfield  {journal} {\bibinfo
  {journal} {\prd}\ }\textbf {\bibinfo {volume} {93}},\ \bibinfo {eid} {123003}
  (\bibinfo {year} {2016})},\ \Eprint {http://arxiv.org/abs/1601.01970}
  {arXiv:1601.01970} \BibitemShut {NoStop}%
\bibitem [{\citenamefont {{Dai}}\ \emph {et~al.}(2013)\citenamefont {{Dai}},
  \citenamefont {{Jeong}}, \citenamefont {{Kamionkowski}},\ and\ \citenamefont
  {{Chluba}}}]{dai}%
  \BibitemOpen
  \bibfield  {author} {\bibinfo {author} {\bibfnamefont {L.}~\bibnamefont
  {{Dai}}}, \bibinfo {author} {\bibfnamefont {D.}~\bibnamefont {{Jeong}}},
  \bibinfo {author} {\bibfnamefont {M.}~\bibnamefont {{Kamionkowski}}}, \ and\
  \bibinfo {author} {\bibfnamefont {J.}~\bibnamefont {{Chluba}}},\ }\bibfield
  {title} {\enquote {\bibinfo {title} {{The pesky power asymmetry}},}\ }\href
  {\doibase 10.1103/PhysRevD.87.123005} {\bibfield  {journal} {\bibinfo
  {journal} {\prd}\ }\textbf {\bibinfo {volume} {87}},\ \bibinfo {eid} {123005}
  (\bibinfo {year} {2013})},\ \Eprint {http://arxiv.org/abs/1303.6949}
  {arXiv:1303.6949 [astro-ph.CO]} \BibitemShut {NoStop}%
\bibitem [{\citenamefont {{Hirata}}(2009)}]{hirata09}%
  \BibitemOpen
  \bibfield  {author} {\bibinfo {author} {\bibfnamefont {C.~M.}\ \bibnamefont
  {{Hirata}}},\ }\bibfield  {title} {\enquote {\bibinfo {title} {{Constraints
  on cosmic hemispherical power anomalies from quasars}},}\ }\href {\doibase
  10.1088/1475-7516/2009/09/011} {\bibfield  {journal} {\bibinfo  {journal}
  {\jcap}\ }\textbf {\bibinfo {volume} {9}},\ \bibinfo {eid} {011} (\bibinfo
  {year} {2009})},\ \Eprint {http://arxiv.org/abs/0907.0703} {arXiv:0907.0703
  [astro-ph.CO]} \BibitemShut {NoStop}%
\bibitem [{\citenamefont {{Hanson}}\ and\ \citenamefont
  {{Lewis}}(2009)}]{hansonlewis}%
  \BibitemOpen
  \bibfield  {author} {\bibinfo {author} {\bibfnamefont {D.}~\bibnamefont
  {{Hanson}}}\ and\ \bibinfo {author} {\bibfnamefont {A.}~\bibnamefont
  {{Lewis}}},\ }\bibfield  {title} {\enquote {\bibinfo {title} {{Estimators for
  CMB statistical anisotropy}},}\ }\href {\doibase 10.1103/PhysRevD.80.063004}
  {\bibfield  {journal} {\bibinfo  {journal} {\prd}\ }\textbf {\bibinfo
  {volume} {80}},\ \bibinfo {eid} {063004} (\bibinfo {year} {2009})},\ \Eprint
  {http://arxiv.org/abs/0908.0963} {arXiv:0908.0963 [astro-ph.CO]} \BibitemShut
  {NoStop}%
\bibitem [{\citenamefont {{Rybicki}}\ and\ \citenamefont
  {{Press}}(1992)}]{rybickipress}%
  \BibitemOpen
  \bibfield  {author} {\bibinfo {author} {\bibfnamefont {G.~B.}\ \bibnamefont
  {{Rybicki}}}\ and\ \bibinfo {author} {\bibfnamefont {W.~H.}\ \bibnamefont
  {{Press}}},\ }\bibfield  {title} {\enquote {\bibinfo {title} {{Interpolation,
  realization, and reconstruction of noisy, irregularly sampled data}},}\
  }\href {\doibase 10.1086/171845} {\bibfield  {journal} {\bibinfo  {journal}
  {\apj}\ }\textbf {\bibinfo {volume} {398}},\ \bibinfo {pages} {169--176}
  (\bibinfo {year} {1992})}\BibitemShut {NoStop}%
\bibitem [{\citenamefont {{Hoffman}}\ and\ \citenamefont
  {{Ribak}}(1991)}]{hoffman91}%
  \BibitemOpen
  \bibfield  {author} {\bibinfo {author} {\bibfnamefont {Y.}~\bibnamefont
  {{Hoffman}}}\ and\ \bibinfo {author} {\bibfnamefont {E.}~\bibnamefont
  {{Ribak}}},\ }\bibfield  {title} {\enquote {\bibinfo {title} {{Constrained
  realizations of Gaussian fields - A simple algorithm}},}\ }\href {\doibase
  10.1086/186160} {\bibfield  {journal} {\bibinfo  {journal} {\apjl}\ }\textbf
  {\bibinfo {volume} {380}},\ \bibinfo {pages} {L5--L8} (\bibinfo {year}
  {1991})}\BibitemShut {NoStop}%
\bibitem [{\citenamefont {{Hoffman}}\ and\ \citenamefont
  {{Ribak}}(1992)}]{hoffman92}%
  \BibitemOpen
  \bibfield  {author} {\bibinfo {author} {\bibfnamefont {Y.}~\bibnamefont
  {{Hoffman}}}\ and\ \bibinfo {author} {\bibfnamefont {E.}~\bibnamefont
  {{Ribak}}},\ }\bibfield  {title} {\enquote {\bibinfo {title} {{Primordial
  Gaussian perturbation fields - Constrained realizations}},}\ }\href {\doibase
  10.1086/170886} {\bibfield  {journal} {\bibinfo  {journal} {\apj}\ }\textbf
  {\bibinfo {volume} {384}},\ \bibinfo {pages} {448--452} (\bibinfo {year}
  {1992})}\BibitemShut {NoStop}%
\bibitem [{\citenamefont {{Bunn}}\ \emph {et~al.}(1994)\citenamefont {{Bunn}},
  \citenamefont {{Fisher}}, \citenamefont {{Hoffman}}, \citenamefont {{Lahav}},
  \citenamefont {{Silk}},\ and\ \citenamefont {{Zaroubi}}}]{bunnwiener}%
  \BibitemOpen
  \bibfield  {author} {\bibinfo {author} {\bibfnamefont {E.~F.}\ \bibnamefont
  {{Bunn}}}, \bibinfo {author} {\bibfnamefont {K.~B.}\ \bibnamefont
  {{Fisher}}}, \bibinfo {author} {\bibfnamefont {Y.}~\bibnamefont {{Hoffman}}},
  \bibinfo {author} {\bibfnamefont {O.}~\bibnamefont {{Lahav}}}, \bibinfo
  {author} {\bibfnamefont {J.}~\bibnamefont {{Silk}}}, \ and\ \bibinfo {author}
  {\bibfnamefont {S.}~\bibnamefont {{Zaroubi}}},\ }\bibfield  {title} {\enquote
  {\bibinfo {title} {{Wiener filtering of the COBE Differential Microwave
  Radiometer data}},}\ }\href {\doibase 10.1086/187515} {\bibfield  {journal}
  {\bibinfo  {journal} {\apjl}\ }\textbf {\bibinfo {volume} {432}},\ \bibinfo
  {pages} {L75--L78} (\bibinfo {year} {1994})},\ \Eprint
  {http://arxiv.org/abs/astro-ph/9404007} {astro-ph/9404007} \BibitemShut
  {NoStop}%
\bibitem [{\citenamefont {{Lahav}}\ \emph {et~al.}(1994)\citenamefont
  {{Lahav}}, \citenamefont {{Fisher}}, \citenamefont {{Hoffman}}, \citenamefont
  {{Scharf}},\ and\ \citenamefont {{Zaroubi}}}]{lahav}%
  \BibitemOpen
  \bibfield  {author} {\bibinfo {author} {\bibfnamefont {O.}~\bibnamefont
  {{Lahav}}}, \bibinfo {author} {\bibfnamefont {K.~B.}\ \bibnamefont
  {{Fisher}}}, \bibinfo {author} {\bibfnamefont {Y.}~\bibnamefont {{Hoffman}}},
  \bibinfo {author} {\bibfnamefont {C.~A.}\ \bibnamefont {{Scharf}}}, \ and\
  \bibinfo {author} {\bibfnamefont {S.}~\bibnamefont {{Zaroubi}}},\ }\bibfield
  {title} {\enquote {\bibinfo {title} {{Wiener Reconstruction of All-Sky Galaxy
  Surveys in Spherical Harmonics}},}\ }\href {\doibase 10.1086/187244}
  {\bibfield  {journal} {\bibinfo  {journal} {\apjl}\ }\textbf {\bibinfo
  {volume} {423}},\ \bibinfo {pages} {L93} (\bibinfo {year} {1994})},\ \Eprint
  {http://arxiv.org/abs/astro-ph/9311059} {astro-ph/9311059} \BibitemShut
  {NoStop}%
\bibitem [{\citenamefont {{Bunn}}\ \emph {et~al.}(1996)\citenamefont {{Bunn}},
  \citenamefont {{Hoffman}},\ and\ \citenamefont {{Silk}}}]{bunnwiener2}%
  \BibitemOpen
  \bibfield  {author} {\bibinfo {author} {\bibfnamefont {E.~F.}\ \bibnamefont
  {{Bunn}}}, \bibinfo {author} {\bibfnamefont {Y.}~\bibnamefont {{Hoffman}}}, \
  and\ \bibinfo {author} {\bibfnamefont {J.}~\bibnamefont {{Silk}}},\
  }\bibfield  {title} {\enquote {\bibinfo {title} {{The Wiener-filtered COBE
  DMR Data and Predictions for the Tenerife Experiment}},}\ }\href {\doibase
  10.1086/177294} {\bibfield  {journal} {\bibinfo  {journal} {\apj}\ }\textbf
  {\bibinfo {volume} {464}},\ \bibinfo {pages} {1} (\bibinfo {year} {1996})},\
  \Eprint {http://arxiv.org/abs/astro-ph/9509045} {astro-ph/9509045}
  \BibitemShut {NoStop}%
\bibitem [{\citenamefont {{Karhunen}}(1947)}]{karhunen}%
  \BibitemOpen
  \bibfield  {author} {\bibinfo {author} {\bibfnamefont {K.}~\bibnamefont
  {{Karhunen}}},\ }\bibfield  {title} {\enquote {\bibinfo {title} {{\"Uber
  lineare Methoden in der Wahrscheinlichkeitsrechnung}},}\ }\href@noop {}
  {\bibfield  {journal} {\bibinfo  {journal} {Ann.\ Acad. Sci. Fennicae Ser.\
  A.\ I.\ Math.-Phys.}\ }\textbf {\bibinfo {volume} {37}},\ \bibinfo {pages}
  {1--79} (\bibinfo {year} {1947})}\BibitemShut {NoStop}%
\bibitem [{\citenamefont {{Bond}}(1995)}]{bondkl}%
  \BibitemOpen
  \bibfield  {author} {\bibinfo {author} {\bibfnamefont {J.~R.}\ \bibnamefont
  {{Bond}}},\ }\bibfield  {title} {\enquote {\bibinfo {title} {{Signal-to-Noise
  Eigenmode Analysis of the Two-Year COBE Maps}},}\ }\href {\doibase
  10.1103/PhysRevLett.74.4369} {\bibfield  {journal} {\bibinfo  {journal}
  {Physical Review Letters}\ }\textbf {\bibinfo {volume} {74}},\ \bibinfo
  {pages} {4369--4372} (\bibinfo {year} {1995})},\ \Eprint
  {http://arxiv.org/abs/astro-ph/9407044} {astro-ph/9407044} \BibitemShut
  {NoStop}%
\bibitem [{\citenamefont {{Bunn}}\ \emph {et~al.}(1995)\citenamefont {{Bunn}},
  \citenamefont {{Scott}},\ and\ \citenamefont {{White}}}]{bunnscottwhite}%
  \BibitemOpen
  \bibfield  {author} {\bibinfo {author} {\bibfnamefont {E.~F.}\ \bibnamefont
  {{Bunn}}}, \bibinfo {author} {\bibfnamefont {D.}~\bibnamefont {{Scott}}}, \
  and\ \bibinfo {author} {\bibfnamefont {M.}~\bibnamefont {{White}}},\
  }\bibfield  {title} {\enquote {\bibinfo {title} {{The COBE normalization for
  standard cold dark matter}},}\ }\href {\doibase 10.1086/187776} {\bibfield
  {journal} {\bibinfo  {journal} {\apjl}\ }\textbf {\bibinfo {volume} {441}},\
  \bibinfo {pages} {L9--L12} (\bibinfo {year} {1995})},\ \Eprint
  {http://arxiv.org/abs/astro-ph/9409003} {astro-ph/9409003} \BibitemShut
  {NoStop}%
\bibitem [{\citenamefont {{Bunn}}\ and\ \citenamefont
  {{Sugiyama}}(1995)}]{bunnsugiyama}%
  \BibitemOpen
  \bibfield  {author} {\bibinfo {author} {\bibfnamefont {E.~F.}\ \bibnamefont
  {{Bunn}}}\ and\ \bibinfo {author} {\bibfnamefont {N.}~\bibnamefont
  {{Sugiyama}}},\ }\bibfield  {title} {\enquote {\bibinfo {title}
  {{Cosmological Constant Cold Dark Matter Models and the COBE Two-Year Sky
  Maps}},}\ }\href {\doibase 10.1086/175765} {\bibfield  {journal} {\bibinfo
  {journal} {\apj}\ }\textbf {\bibinfo {volume} {446}},\ \bibinfo {pages} {49}
  (\bibinfo {year} {1995})},\ \Eprint {http://arxiv.org/abs/astro-ph/9407069}
  {astro-ph/9407069} \BibitemShut {NoStop}%
\bibitem [{\citenamefont {{Vogeley}}(1995)}]{vogeley}%
  \BibitemOpen
  \bibfield  {author} {\bibinfo {author} {\bibfnamefont {M.}~\bibnamefont
  {{Vogeley}}},\ }\bibfield  {title} {\enquote {\bibinfo {title} {{Constraints
  on Cosmological Models from Once and Future Surveys}},}\ }in\ \href@noop {}
  {\emph {\bibinfo {booktitle} {Wide Field Spectroscopy and the Distant
  Universe}}},\ \bibinfo {editor} {edited by\ \bibinfo {editor} {\bibfnamefont
  {S.~J.}\ \bibnamefont {{Maddox}}}\ and\ \bibinfo {editor} {\bibfnamefont
  {A.}~\bibnamefont {{Aragon-Salamanca}}}}\ (\bibinfo {year} {1995})\ p.\
  \bibinfo {pages} {142},\ \Eprint {http://arxiv.org/abs/astro-ph/9410068}
  {astro-ph/9410068} \BibitemShut {NoStop}%
\bibitem [{\citenamefont {{Vogeley}}\ and\ \citenamefont
  {{Szalay}}(1996)}]{vogeleyszalay}%
  \BibitemOpen
  \bibfield  {author} {\bibinfo {author} {\bibfnamefont {M.~S.}\ \bibnamefont
  {{Vogeley}}}\ and\ \bibinfo {author} {\bibfnamefont {A.~S.}\ \bibnamefont
  {{Szalay}}},\ }\bibfield  {title} {\enquote {\bibinfo {title} {{Eigenmode
  Analysis of Galaxy Redshift Surveys. I. Theory and Methods}},}\ }\href
  {\doibase 10.1086/177399} {\bibfield  {journal} {\bibinfo  {journal} {\apj}\
  }\textbf {\bibinfo {volume} {465}},\ \bibinfo {pages} {34} (\bibinfo {year}
  {1996})},\ \Eprint {http://arxiv.org/abs/astro-ph/9601185} {astro-ph/9601185}
  \BibitemShut {NoStop}%
\bibitem [{\citenamefont {{G{\'o}rski}}\ \emph {et~al.}(2005)\citenamefont
  {{G{\'o}rski}}, \citenamefont {{Hivon}}, \citenamefont {{Banday}},
  \citenamefont {{Wandelt}}, \citenamefont {{Hansen}}, \citenamefont
  {{Reinecke}},\ and\ \citenamefont {{Bartelmann}}}]{healpix}%
  \BibitemOpen
  \bibfield  {author} {\bibinfo {author} {\bibfnamefont {K.~M.}\ \bibnamefont
  {{G{\'o}rski}}}, \bibinfo {author} {\bibfnamefont {E.}~\bibnamefont
  {{Hivon}}}, \bibinfo {author} {\bibfnamefont {A.~J.}\ \bibnamefont
  {{Banday}}}, \bibinfo {author} {\bibfnamefont {B.~D.}\ \bibnamefont
  {{Wandelt}}}, \bibinfo {author} {\bibfnamefont {F.~K.}\ \bibnamefont
  {{Hansen}}}, \bibinfo {author} {\bibfnamefont {M.}~\bibnamefont
  {{Reinecke}}}, \ and\ \bibinfo {author} {\bibfnamefont {M.}~\bibnamefont
  {{Bartelmann}}},\ }\bibfield  {title} {\enquote {\bibinfo {title} {{HEALPix:
  A Framework for High-Resolution Discretization and Fast Analysis of Data
  Distributed on the Sphere}},}\ }\href {\doibase 10.1086/427976} {\bibfield
  {journal} {\bibinfo  {journal} {\apj}\ }\textbf {\bibinfo {volume} {622}},\
  \bibinfo {pages} {759--771} (\bibinfo {year} {2005})},\ \Eprint
  {http://arxiv.org/abs/astro-ph/0409513} {astro-ph/0409513} \BibitemShut
  {NoStop}%
\bibitem [{\citenamefont {{Abazajian}}\ \emph {et~al.}(2016)\citenamefont
  {{Abazajian}}, \citenamefont {{Adshead}}, \citenamefont {{Ahmed}},
  \citenamefont {{Allen}}, \citenamefont {{Alonso}}, \citenamefont {{Arnold}},
  \citenamefont {{Baccigalupi}}, \citenamefont {{Bartlett}}, \citenamefont
  {{Battaglia}}, \citenamefont {{Benson}}, \citenamefont {{Bischoff}},
  \citenamefont {{Borrill}}, \citenamefont {{Buza}}, \citenamefont
  {{Calabrese}}, \citenamefont {{Caldwell}}, \citenamefont {{Carlstrom}},
  \citenamefont {{Chang}}, \citenamefont {{Crawford}}, \citenamefont
  {{Cyr-Racine}}, \citenamefont {{De Bernardis}}, \citenamefont {{de Haan}},
  \citenamefont {{di Serego Alighieri}}, \citenamefont {{Dunkley}},
  \citenamefont {{Dvorkin}}, \citenamefont {{Errard}}, \citenamefont
  {{Fabbian}}, \citenamefont {{Feeney}}, \citenamefont {{Ferraro}},
  \citenamefont {{Filippini}}, \citenamefont {{Flauger}}, \citenamefont
  {{Fuller}}, \citenamefont {{Gluscevic}}, \citenamefont {{Green}},
  \citenamefont {{Grin}}, \citenamefont {{Grohs}}, \citenamefont {{Henning}},
  \citenamefont {{Hill}}, \citenamefont {{Hlozek}}, \citenamefont {{Holder}},
  \citenamefont {{Holzapfel}}, \citenamefont {{Hu}}, \citenamefont
  {{Huffenberger}}, \citenamefont {{Keskitalo}}, \citenamefont {{Knox}},
  \citenamefont {{Kosowsky}}, \citenamefont {{Kovac}}, \citenamefont
  {{Kovetz}}, \citenamefont {{Kuo}}, \citenamefont {{Kusaka}}, \citenamefont
  {{Le Jeune}}, \citenamefont {{Lee}}, \citenamefont {{Lilley}}, \citenamefont
  {{Loverde}}, \citenamefont {{Madhavacheril}}, \citenamefont {{Mantz}},
  \citenamefont {{Marsh}}, \citenamefont {{McMahon}}, \citenamefont
  {{Meerburg}}, \citenamefont {{Meyers}}, \citenamefont {{Miller}},
  \citenamefont {{Munoz}}, \citenamefont {{Nguyen}}, \citenamefont {{Niemack}},
  \citenamefont {{Peloso}}, \citenamefont {{Peloton}}, \citenamefont
  {{Pogosian}}, \citenamefont {{Pryke}}, \citenamefont {{Raveri}},
  \citenamefont {{Reichardt}}, \citenamefont {{Rocha}}, \citenamefont
  {{Rotti}}, \citenamefont {{Schaan}}, \citenamefont {{Schmittfull}},
  \citenamefont {{Scott}}, \citenamefont {{Sehgal}}, \citenamefont
  {{Shandera}}, \citenamefont {{Sherwin}}, \citenamefont {{Smith}},
  \citenamefont {{Sorbo}}, \citenamefont {{Starkman}}, \citenamefont {{Story}},
  \citenamefont {{van Engelen}}, \citenamefont {{Vieira}}, \citenamefont
  {{Watson}}, \citenamefont {{Whitehorn}},\ and\ \citenamefont {{Kimmy
  Wu}}}]{s4}%
  \BibitemOpen
  \bibfield  {author} {\bibinfo {author} {\bibfnamefont {K.~N.}\ \bibnamefont
  {{Abazajian}}}, \bibinfo {author} {\bibfnamefont {P.}~\bibnamefont
  {{Adshead}}}, \bibinfo {author} {\bibfnamefont {Z.}~\bibnamefont {{Ahmed}}},
  \bibinfo {author} {\bibfnamefont {S.~W.}\ \bibnamefont {{Allen}}}, \bibinfo
  {author} {\bibfnamefont {D.}~\bibnamefont {{Alonso}}}, \bibinfo {author}
  {\bibfnamefont {K.~S.}\ \bibnamefont {{Arnold}}}, \bibinfo {author}
  {\bibfnamefont {C.}~\bibnamefont {{Baccigalupi}}}, \bibinfo {author}
  {\bibfnamefont {J.~G.}\ \bibnamefont {{Bartlett}}}, \bibinfo {author}
  {\bibfnamefont {N.}~\bibnamefont {{Battaglia}}}, \bibinfo {author}
  {\bibfnamefont {B.~A.}\ \bibnamefont {{Benson}}}, \bibinfo {author}
  {\bibfnamefont {C.~A.}\ \bibnamefont {{Bischoff}}}, \bibinfo {author}
  {\bibfnamefont {J.}~\bibnamefont {{Borrill}}}, \bibinfo {author}
  {\bibfnamefont {V.}~\bibnamefont {{Buza}}}, \bibinfo {author} {\bibfnamefont
  {E.}~\bibnamefont {{Calabrese}}}, \bibinfo {author} {\bibfnamefont
  {R.}~\bibnamefont {{Caldwell}}}, \bibinfo {author} {\bibfnamefont {J.~E.}\
  \bibnamefont {{Carlstrom}}}, \bibinfo {author} {\bibfnamefont {C.~L.}\
  \bibnamefont {{Chang}}}, \bibinfo {author} {\bibfnamefont {T.~M.}\
  \bibnamefont {{Crawford}}}, \bibinfo {author} {\bibfnamefont {F.-Y.}\
  \bibnamefont {{Cyr-Racine}}}, \bibinfo {author} {\bibfnamefont
  {F.}~\bibnamefont {{De Bernardis}}}, \bibinfo {author} {\bibfnamefont
  {T.}~\bibnamefont {{de Haan}}}, \bibinfo {author} {\bibfnamefont
  {S.}~\bibnamefont {{di Serego Alighieri}}}, \bibinfo {author} {\bibfnamefont
  {J.}~\bibnamefont {{Dunkley}}}, \bibinfo {author} {\bibfnamefont
  {C.}~\bibnamefont {{Dvorkin}}}, \bibinfo {author} {\bibfnamefont
  {J.}~\bibnamefont {{Errard}}}, \bibinfo {author} {\bibfnamefont
  {G.}~\bibnamefont {{Fabbian}}}, \bibinfo {author} {\bibfnamefont
  {S.}~\bibnamefont {{Feeney}}}, \bibinfo {author} {\bibfnamefont
  {S.}~\bibnamefont {{Ferraro}}}, \bibinfo {author} {\bibfnamefont {J.~P.}\
  \bibnamefont {{Filippini}}}, \bibinfo {author} {\bibfnamefont
  {R.}~\bibnamefont {{Flauger}}}, \bibinfo {author} {\bibfnamefont {G.~M.}\
  \bibnamefont {{Fuller}}}, \bibinfo {author} {\bibfnamefont {V.}~\bibnamefont
  {{Gluscevic}}}, \bibinfo {author} {\bibfnamefont {D.}~\bibnamefont
  {{Green}}}, \bibinfo {author} {\bibfnamefont {D.}~\bibnamefont {{Grin}}},
  \bibinfo {author} {\bibfnamefont {E.}~\bibnamefont {{Grohs}}}, \bibinfo
  {author} {\bibfnamefont {J.~W.}\ \bibnamefont {{Henning}}}, \bibinfo {author}
  {\bibfnamefont {J.~C.}\ \bibnamefont {{Hill}}}, \bibinfo {author}
  {\bibfnamefont {R.}~\bibnamefont {{Hlozek}}}, \bibinfo {author}
  {\bibfnamefont {G.}~\bibnamefont {{Holder}}}, \bibinfo {author}
  {\bibfnamefont {W.}~\bibnamefont {{Holzapfel}}}, \bibinfo {author}
  {\bibfnamefont {W.}~\bibnamefont {{Hu}}}, \bibinfo {author} {\bibfnamefont
  {K.~M.}\ \bibnamefont {{Huffenberger}}}, \bibinfo {author} {\bibfnamefont
  {R.}~\bibnamefont {{Keskitalo}}}, \bibinfo {author} {\bibfnamefont
  {L.}~\bibnamefont {{Knox}}}, \bibinfo {author} {\bibfnamefont
  {A.}~\bibnamefont {{Kosowsky}}}, \bibinfo {author} {\bibfnamefont
  {J.}~\bibnamefont {{Kovac}}}, \bibinfo {author} {\bibfnamefont {E.~D.}\
  \bibnamefont {{Kovetz}}}, \bibinfo {author} {\bibfnamefont {C.-L.}\
  \bibnamefont {{Kuo}}}, \bibinfo {author} {\bibfnamefont {A.}~\bibnamefont
  {{Kusaka}}}, \bibinfo {author} {\bibfnamefont {M.}~\bibnamefont {{Le
  Jeune}}}, \bibinfo {author} {\bibfnamefont {A.~T.}\ \bibnamefont {{Lee}}},
  \bibinfo {author} {\bibfnamefont {M.}~\bibnamefont {{Lilley}}}, \bibinfo
  {author} {\bibfnamefont {M.}~\bibnamefont {{Loverde}}}, \bibinfo {author}
  {\bibfnamefont {M.~S.}\ \bibnamefont {{Madhavacheril}}}, \bibinfo {author}
  {\bibfnamefont {A.}~\bibnamefont {{Mantz}}}, \bibinfo {author} {\bibfnamefont
  {D.~J.~E.}\ \bibnamefont {{Marsh}}}, \bibinfo {author} {\bibfnamefont
  {J.}~\bibnamefont {{McMahon}}}, \bibinfo {author} {\bibfnamefont {P.~D.}\
  \bibnamefont {{Meerburg}}}, \bibinfo {author} {\bibfnamefont
  {J.}~\bibnamefont {{Meyers}}}, \bibinfo {author} {\bibfnamefont {A.~D.}\
  \bibnamefont {{Miller}}}, \bibinfo {author} {\bibfnamefont {J.~B.}\
  \bibnamefont {{Munoz}}}, \bibinfo {author} {\bibfnamefont {H.~N.}\
  \bibnamefont {{Nguyen}}}, \bibinfo {author} {\bibfnamefont {M.~D.}\
  \bibnamefont {{Niemack}}}, \bibinfo {author} {\bibfnamefont {M.}~\bibnamefont
  {{Peloso}}}, \bibinfo {author} {\bibfnamefont {J.}~\bibnamefont {{Peloton}}},
  \bibinfo {author} {\bibfnamefont {L.}~\bibnamefont {{Pogosian}}}, \bibinfo
  {author} {\bibfnamefont {C.}~\bibnamefont {{Pryke}}}, \bibinfo {author}
  {\bibfnamefont {M.}~\bibnamefont {{Raveri}}}, \bibinfo {author}
  {\bibfnamefont {C.~L.}\ \bibnamefont {{Reichardt}}}, \bibinfo {author}
  {\bibfnamefont {G.}~\bibnamefont {{Rocha}}}, \bibinfo {author} {\bibfnamefont
  {A.}~\bibnamefont {{Rotti}}}, \bibinfo {author} {\bibfnamefont
  {E.}~\bibnamefont {{Schaan}}}, \bibinfo {author} {\bibfnamefont {M.~M.}\
  \bibnamefont {{Schmittfull}}}, \bibinfo {author} {\bibfnamefont
  {D.}~\bibnamefont {{Scott}}}, \bibinfo {author} {\bibfnamefont
  {N.}~\bibnamefont {{Sehgal}}}, \bibinfo {author} {\bibfnamefont
  {S.}~\bibnamefont {{Shandera}}}, \bibinfo {author} {\bibfnamefont {B.~D.}\
  \bibnamefont {{Sherwin}}}, \bibinfo {author} {\bibfnamefont {T.~L.}\
  \bibnamefont {{Smith}}}, \bibinfo {author} {\bibfnamefont {L.}~\bibnamefont
  {{Sorbo}}}, \bibinfo {author} {\bibfnamefont {G.~D.}\ \bibnamefont
  {{Starkman}}}, \bibinfo {author} {\bibfnamefont {K.~T.}\ \bibnamefont
  {{Story}}}, \bibinfo {author} {\bibfnamefont {A.}~\bibnamefont {{van
  Engelen}}}, \bibinfo {author} {\bibfnamefont {J.~D.}\ \bibnamefont
  {{Vieira}}}, \bibinfo {author} {\bibfnamefont {S.}~\bibnamefont {{Watson}}},
  \bibinfo {author} {\bibfnamefont {N.}~\bibnamefont {{Whitehorn}}}, \ and\
  \bibinfo {author} {\bibfnamefont {W.~L.}\ \bibnamefont {{Kimmy Wu}}},\
  }\bibfield  {title} {\enquote {\bibinfo {title} {{CMB-S4 Science Book, First
  Edition}},}\ }\href@noop {} {\bibfield  {journal} {\bibinfo  {journal} {ArXiv
  e-prints}\ } (\bibinfo {year} {2016})},\ \Eprint
  {http://arxiv.org/abs/1610.02743} {arXiv:1610.02743} \BibitemShut {NoStop}%
\end{thebibliography}%

\end{document}